\documentclass[aps,prl,notitlepage,superscriptaddress,showpacs,twocolumn]{revtex4-1}
\usepackage{graphicx,subfigure,epsfig}
 \usepackage{array,multirow}
\usepackage{dcolumn}
\usepackage{amssymb,amsmath,amsfonts,mathrsfs}
\usepackage{array}
\usepackage{times,setspace}
\usepackage{latexsym}
\usepackage{float,flafter,bm,bbm}
\usepackage{epstopdf,color,multirow}
\usepackage[colorlinks,linkcolor=blue,anchorcolor=blue,urlcolor=blue,citecolor=blue]{hyperref}
\usepackage{footnote}
\usepackage{booktabs}
\usepackage{natbib}
\hypersetup{
    colorlinks=true,
    linkcolor=blue,
    filecolor=magenta,
    urlcolor=blue,
}
\begin{document}

\title{Quantum Information Geometry of Multicomponent Superconducting Fluctuation Transport}

\author{Zi-Ting Sun}\thanks{zsunaw@connect.ust.hk}

\author{Ying-Ming Xie}\thanks{yxieai@connect.ust.hk}
\affiliation{RIKEN Center for Emergent Matter Science (CEMS), Wako, Saitama 351-0198, Japan} 	
\author{Naoto Nagaosa}\thanks{nagaosa@riken.jp}
\affiliation{RIKEN Center for Emergent Matter Science (CEMS), Wako, Saitama 351-0198, Japan} 	
   
 \affiliation{Fundamental Quantum Science Program (FQSP), TRIP Headquarters, RIKEN, Wako 351-0198, Japan}

	\date{\today}
	\begin{abstract}
Quantum geometry underlies many electronic responses, but its transport signatures have so far been established mainly for pure single-particle Bloch states. Whether collective many-body fluctuations possess a measurable quantum geometry remains largely unexplored. Here we show that superconducting fluctuation transport provides a direct probe of quantum information geometry in collective many-body matter. Starting from a multicomponent time-dependent Ginzburg–Landau theory in the Gaussian fluctuation regime, we identify the equilibrium density matrix of fluctuating Cooper pairs as the static pair propagator, which defines a positive mixed-state manifold in momentum space. The geometry of this manifold is directly measurable through paraconductivity: the longitudinal paraconductivity is governed by the quantum Fisher information of superconducting fluctuation modes, while the fluctuational anomalous Hall effect is governed by the mean Uhlmann curvature, the mixed-state counterpart of Berry curvature. This correspondence further yields geometric bounds between these two transport components, with no direct analogue in normal electronic transport. Applied to chiral superconducting fluctuations in quarter-metal systems motivated by rhombohedral multilayer graphene, a symmetry-allowed Lifshitz invariant generates finite mean Uhlmann curvature and logarithmically enhances the anomalous Hall conductivity above $T_c$. Our results establish collective superconducting fluctuations as an experimentally accessible transport probe of mixed-state quantum information geometry.
	\end{abstract}
	\maketitle

\section{Introduction}
Quantum geometry has become a central organizing principle for electronic responses in crystalline quantum matter \cite{gao2025quantum,yu2025quantum,jiang2025revealing,liu2025quantum}. As representative phenomena, Berry curvature and quantum metric enter the anomalous Hall effect~\cite{nagaosa2010anomalous,xiao2010berry,chang2023colloquium}, orbital magnetization~\cite{vanderbilt2018berry}, nonlinear Hall effects~\cite{sodemann2015quantum,gao2014field,wang2021intrinsic,liu2021intrinsic,zhang2023higher}, nonlinear optical responses \cite{ahn2022riemannian} and geometric contributions to superconducting stiffness~\cite{peotta2015superfluidity}. These results show that material responses are controlled not only by spectra and occupations, but also by the geometry of quantum states in parameter space.

Yet these developments have largely focused on pure-state electronic Bloch wave functions, leaving open whether collective many-body fluctuations can carry a quantum geometry that is accessible through experiment. For fluctuations near finite-temperature phase transitions, the carriers are thermally populated many-body modes and are therefore
intrinsically mixed-state objects. Their natural geometric descriptor is quantum information geometry, which extends the pure-state quantum geometric tensor to mixed states~\cite{uhlmann1986parallel,uhlmann1991gauge,braunstein1994statistical,carollo2018uhlmann,carollo2020geometry}. Although these concepts are well developed in quantum metrology, finite-temperature topological phases, open quantum systems, and optical sum rules~\cite{albert2016geometry,leonforte2019uhlmann,liu2020quantum,ji2025density}, their connection to measurable transport of collective modes remains largely unexplored.

\begin{figure}
		\centering
		\includegraphics[width=1\linewidth]{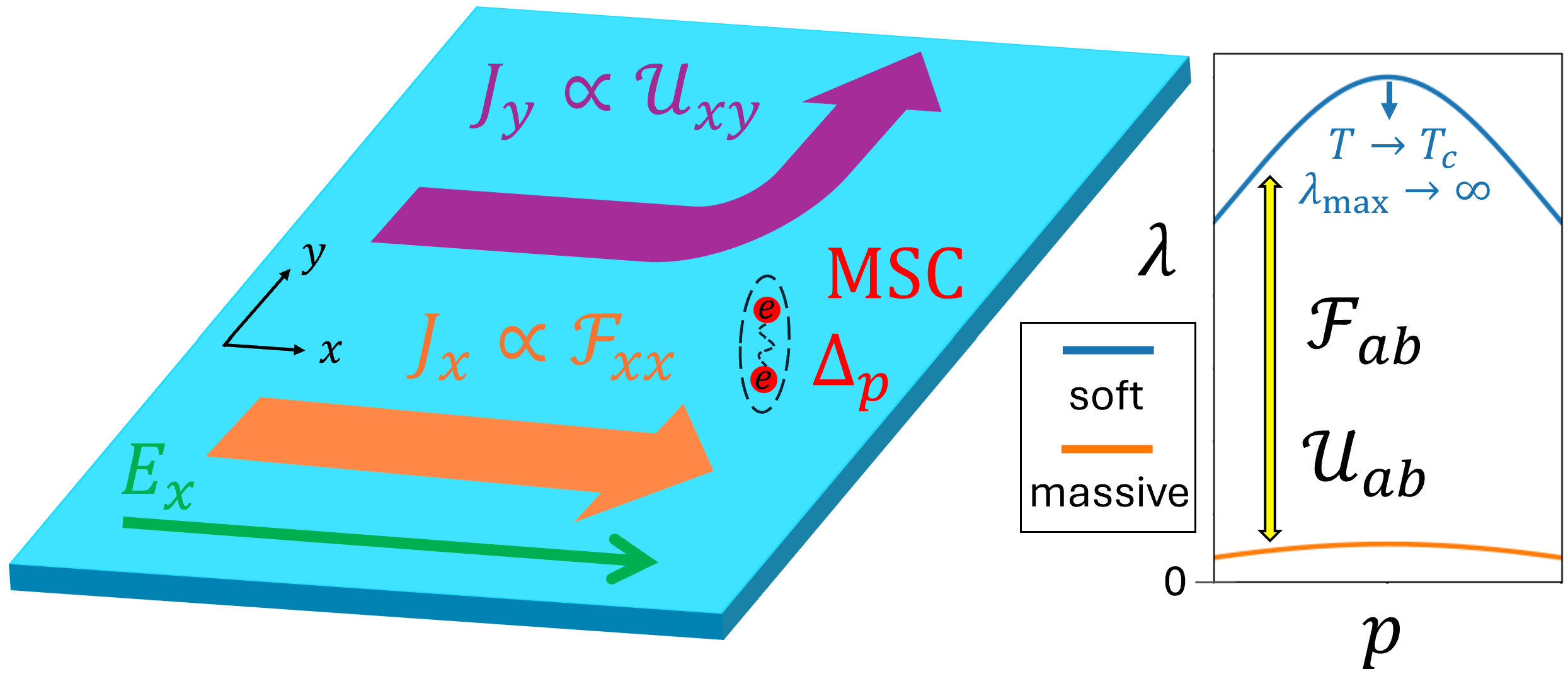}
		\caption{\textbf{Schematic illustration of superconductor fluctuation transport as a probe of quantum information geometry.} Left panel: An applied electric field \(E_x\) drives fluctuating Cooper pairs $\Delta_{\mathbf{p}}$ of a multicomponent superconductor (MSC) in momentum space. The longitudinal paraconductivity is governed by the quantum Fisher information, giving \(J_x\propto \mathcal F_{xx}\), while the transverse anomalous Hall paraconductivity is governed by the mean Uhlmann curvature, giving \(J_y\propto \mathcal U_{xy}\). The right panel shows the spectrum of the pair-propagator eigenvalues \(\lambda\): upon approaching the superconducting transition, \(T\to T_c\), the softest fluctuation mode diverges, \(\lambda_{\max}\to\infty\), whereas the massive mode remains finite. The mixed-state geometry emerges due to the variation of pairing eigen-functions in momentum space.}
		\label{fig:fig1}
\end{figure}

Superconducting fluctuations above the critical temperature $T_c$ provide a concrete route to this problem. Paraconductivity is carried by short-lived fluctuating Cooper pairs~\cite{schmid1969diamagnetic,larkin2005theory}, which in multicomponent superconductors may also carry orbital, valley, spin, or chiral internal degrees of freedom. The pair propagator then forms a momentum-space family of positive matrices whose quantum information geometry can enter fluctuation conductivity. Theories of superconducting fluctuation transport are well established~\cite{fukuyama1971fluctuation,aronov1995gauge,michaeli2012hall,sumiyoshi2014giant,li2020fluctuational}, however, what remains missing is a transport principle connecting measurable fluctuation currents to the mixed-state geometry of collective Cooper-pair modes.

In this work, we reveal that the superconducting fluctuation transport can be organized by the quantum information geometry of the pair-propagator manifold. We develop this framework starting from a multicomponent time-dependent Ginzburg-Landau (TDGL) equation in the Gaussian fluctuation regime. In this regime, the equilibrium pairing-fluctuation density matrix (PFDM) is identical to the pair propagator and defines a mixed-state manifold over Cooper pair momentum space. Using the symmetric logarithmic derivative, we construct the quantum Fisher tensor (QFI) and the mean Uhlmann curvature (MUC) of this manifold. We then derive a general expression for the linear paraconductivity, in which the longitudinal dissipative paraconductivity is governed by the QFI, whereas the intrinsic fluctuational anomalous Hall conductivity is governed by the MUC, as schematically illustrated in the left panel of Fig.~\ref{fig:fig1}. The latter requires a finite reactive, particle-hole-asymmetric fluctuation dynamics, clarifying the conditions under which a fluctuation Hall response can serve as a transport signature of finite mixed-state curvature. As a consequence, we point out geometric bound relations between these two transport phenomena, with no direct counterpart in normal-state transport.

As an application, we study chiral superconducting fluctuations in quarter-metal systems motivated by rhombohedral multilayer graphene~\cite{han2025signatures,dutta2026reconfigurable,sheekey2026visualizing}. In this setting, a symmetry-allowed Lifshitz invariant generates finite MUC of the fluctuating Cooper-pair manifold and produces an intrinsic fluctuational anomalous Hall effect above $T_c$, which is logarithmically enhanced near the superconducting transition. These results demonstrate that superconducting fluctuation transport can provide an experimentally accessible probe of quantum information geometry in collective many-body matter.

\section{Results}

\subsection{Quantum information geometry of multicomponent superconducting fluctuations}

We start from the generic Gaussian fluctuation action of a multicomponent superconductor \cite{nagaosa2013quantum,Altland_Simons_2023}:
\begin{equation}
    S_{\mathrm{fl}}[\Delta]
    =
    \sum_{P}
    \Delta^\dagger(P)\,
    \hat{\mathcal{L}}^{-1}[P]\,
    \Delta(P).
    \label{eq:Sfl_tdgl_start}
\end{equation}
   Here \(P=(i\Omega_n,\mathbf p)\) denotes a bosonic four-momentum of the $\Delta$ field, $V$ is the system volume, and throughout this work we set \(k_B=\hbar=1\). \(\hat{\mathcal{L}}\) is the propagator of the multicomponent fluctuating pair field, and the hat indicates a matrix structure in the internal pair-component space. Here, by a multicomponent superconductor we do not simply mean a superconducting material with several microscopic degrees of freedom. Rather, we use the term in an effective low-energy sense, i.e., more than one pairing component remains active because of similar critical temperatures, and the fluctuating order parameter $\Delta$ must be treated as a vector. For example, such components may correspond to sublattice-resolved pairing amplitudes in kagome superconductors
\cite{wu2021nature}, or to symmetry-related chiral pairing channels in unconventional superconductors
\cite{kallin2016chiral}.

   We consider the Gaussian fluctuation region above the superconducting transition, \(T>T_c\). In this region, the higher-order terms in the $\Delta$ field and the higher-frequency $\Omega_n\neq0$ contributions can usually be neglected in the calculations of fluctuation-mediated electromagnetic response. In other words, the classical propagator $\hat L_0(\mathbf p)\equiv\hat{\mathcal{L}}(i\Omega_n=0,\mathbf p)$ is the leading contribution in our scenario. The multicomponent order parameter is also taken as a classical field $\Delta_{\mathbf{p}}$.

We define the eigenmode basis of the pair propagator $\hat L_0$ as \(\hat L_0(\mathbf p)|u_n(\mathbf p)\rangle
    =
    \lambda_n(\mathbf p)|u_n(\mathbf p)\rangle\).
In the Gaussian fluctuation region, all $\lambda_n>0$, thus $\hat L_0$ is positive definite. The critical temperature is determined by \(\max_{n,\mathbf p}
    \lambda_n(\mathbf p,T_c)
    \rightarrow
    \infty\). See the right panel of Fig.~\ref{fig:fig1}. Here we argue that the quantum information geometry of mixed states \cite{braunstein1994statistical} is the most natural description for the quantum geometry of pairing fluctuations. The reason is twofold. First, we are studying a finite-temperature thermal state rather than a zero-temperature ground state. Second, the equilibrium ``density matrix" of the fluctuating order parameter is \(\left\langle\Delta_{\mathbf{p}}\Delta_{\mathbf{p}}^\dagger\right\rangle _{\rm {eq}}=\hat L_0\), so \(\hat L_0\) plays a role analogous to the equilibrium density matrix \cite{ji2025density,guan2026exploring} in the formulation of density matrix geometry.

 Quantum information geometry is the natural extension of pure-state quantum geometry to mixed states. A natural approach to formulate quantum information geometry is based on the symmetric logarithmic differential (SLD) $\hat{I}_a$ \cite{ghosh2026journey} of positive definite $\hat L_0$ \cite{van2022bures} by the Lyapunov equation \(\partial_a \hat L_0
    =\left\{
    \hat L_0,
    \hat{I}_a
    \right\}/2\).
In the eigenbasis of \(\hat L_0\), this gives
\begin{equation}
    (\hat{I}_a)_{mn}
    =\frac{(\partial_a \hat L_0)_{mn}}{\bar{\lambda}_{mn}}=
    \partial_a\ln \lambda_m \delta_{mn}+i
    \frac{\lambda_{mn}}{\bar{\lambda}_{mn}}
    \left(\mathcal A_a\right)_{mn},
    \label{eq:SLD}
\end{equation}
where
\((\partial_a \hat X)_{mn}
    \equiv
    \partial_a (\hat X)_{mn}
    - i [\mathcal A_a,\hat X]_{mn}\)
takes the form of a covariant derivative, \(\left(\mathcal A_a\right)_{mn}=i\langle u_m|\partial_a u_n\rangle\) is the non-Abelian Berry connection of pairing eigenmodes, and we denote $(\hat X)_{mn}\equiv\langle u_m|\hat X| u_n\rangle$. Besides, we write $\bar{\lambda}_{mn}=\left(\lambda_m+\lambda_n\right)/2$ and $\lambda_{mn}=\lambda_m-\lambda_n$ for short. 

In terms of $\hat{I}_a$, we can define the quantum Fisher tensor \cite{carollo2020geometry} of the
pairing fluctuations, \(I_{ab} =
    \mathrm{tr}_{\mathrm{mat}}
    \left[
    \hat L_0
    \hat{I}_a
    \hat{I}_b
    \right]\). \(I_{ab}\) is a Hermitian tensor, whose real (symmetric) part can be identified as the QFI matrix as
\begin{equation}
    \mathcal F_{ab}
    =\operatorname{Re} I_{ab}=
    \frac{1}{2}
    \mathrm{tr}_{\mathrm{mat}}
    \left[
    \hat L_0
    \left\{
    \hat{I}_a,
    \hat{I}_b
    \right\}
    \right],
    \label{eq:unnormalized_QFI_matrix}
\end{equation}
whose spectral form \cite{liu2020quantum} is 
\begin{align}
\mathcal F_{ab}
    &=
    \sum_{m,n}
    \bar \lambda_{mn}
    \left(\hat{I}_a\right)_{nm}
    \left(\hat{I}_b\right)_{mn}\nonumber\\&=
    \sum_n
    \frac{
    (\partial_a\lambda_n)(\partial_b\lambda_n)
    }{
    \lambda_n
    }
    +
    \sum_{m\neq n}
    \frac{
    \lambda_{mn}^2
    }{
    \bar{\lambda}_{mn}
    }
    g^{nm}_{ab}.    
\label{eq:JL_spectral}
\end{align}
Then the corresponding Bures-Wasserstein distance is given by $ds^2_{\rm BW}=\mathcal F_{ab}dp_adp_b/4$ \cite{van2022bures}. Here we used the definition of eigenmode-resolved quantum geometric tensor $Q^{nm}_{ab}=(\mathcal A_a)_{nm}(\mathcal A_b)_{mn}$ and the eigenmode-resolved quantum metric $g^{nm}_{ab}=\operatorname{Re} Q^{nm}_{ab}$. 
The first term in Eq.~\eqref{eq:JL_spectral} is the classical Fisher
information from variations of the eigenvalues \(\lambda_n\), we denote it as $\mathcal F^{\rm cla}_{ab}$.  The second term is the coherent part of the QFI matrix from variations of the eigenvectors \(|u_n\rangle\), and we denote it as $\mathcal F^{\rm coh}_{ab}$. We note that, for single-component superconductors, only $\mathcal F^{\rm cla}_{ab}$ remains. For multi-component superconductors, the nontrivial variation of the eigenvector is related to the coupling between different pairing components at finite momentum $\mathbf p$ in \(\hat L_0(\mathbf p)\) (right panel of Fig.~\ref{fig:fig1}).

For the imaginary (antisymmetric) part of $I_{ab}$, we define the skew-symmetric MUC \cite{uhlmann1986parallel,uhlmann1991gauge,carollo2018uhlmann} as
\begin{equation}
    \mathcal U_{ab}
    =-\frac{1}{2}
    \operatorname{Im} I_{ab}=
    \frac{i}{4}
    \mathrm{tr}_{\mathrm{mat}}
    \left[
    \hat L_0
    [
    \hat{I}_a,
    \hat{I}_b
    ]
    \right].
    \label{eq:unnormalized_MUC_SLD}
\end{equation}
Its spectral representation is
\begin{equation}
    \mathcal U_{ab}
    =
    \frac{i}{4}
    \sum_{m,n}
    \lambda_{nm}
    (\hat{I}_a)_{nm}
    (\hat{I}_b)_{mn}=
    \sum_{n\neq m}
    \frac{
    \lambda_{nm}^{3}
    }{
    \bar\lambda_{mn}^{2}
    }
    \frac{\Omega^{nm}_{ab}}{8}.
    \label{eq:MUC_spectral_Omega}
\end{equation}
where $\Omega^{nm}_{ab}=-2\operatorname{Im}Q^{nm}_{ab}$ is the eigenmode-resolved Berry curvature. Here we note that the quantum Fisher tensor \(I_{ab}\), the QFI $\mathcal F_{ab}$ and $\mathcal U_{ab}$ are the analogues of the quantum geometric tensor, quantum metric and Berry curvature of pure states in mixed states \cite{gao2025quantum}, and are widely used in quantum sensing and quantum metrology. In those cases, the QFI sets the ultimate precision bound for estimating parameters encoded in a quantum state, while the MUC quantifies the quantum incompatibility between different parameter estimations in multiparameter sensing \cite{montenegro2025quantum,ghosh2026journey}. However, as we will see, the quantum information geometry of pairing fluctuations is directly related to the paraconductivity of multicomponent superconductors.

\subsection{Pairing fluctuation density matrix and response currents}


 The standard method to capture the Aslamazov-Larkin contribution to the paraconductivity is through the TDGL equation \cite{ussishkin2002gaussian}. For calculating the paraconductivity, we introduce an electric field in the temporal gauge as $A_0 = 0$, $\mathbf{E}(t) = - \partial_t \mathbf{A}(t)$. Then the order-parameter field  $\Delta_{\mathbf{p}}(t)$ also becomes dynamical. As described in \textbf{Supplementary Materials Note I}~\cite{SM}, the multicomponent TDGL equation \cite{schmid1969diamagnetic} reads:
\begin{equation}
    \Xi \partial_t \Delta_{\mathbf{p}}(t)
    =
    -
    \hat L_0^{-1}
    \bigl(\mathbf{p}-e^*\mathbf{A}(t)\bigr)
    \Delta_{\mathbf{p}}(t)
    +
    \zeta_{\mathbf{p}}(t),
    \label{eq:tdgl_scalar_relaxation}
\end{equation}
where \(e^*=2e\) is the effective charge of the fluctuating Cooper pair, and the gauge covariance implies we can introduce the vector potential through minimal coupling. This equation describes the relaxation of the order parameter configuration $\Delta_{\mathbf{p}}(t)$ towards its equilibrium distribution, interrupted by thermal fluctuations introduced through the white noise $\zeta_{\mathbf{p}}$. Here, we phenomenologically consider a scalar but complex kinetic coefficient \(\Xi =\Xi'+i\Xi''\), with \(\Xi'>0\) and $\hat 1$ is the identity matrix. The real part \(\Xi'\) controls dissipative relaxation, whereas the
imaginary part \(\Xi''\) leads to the precession of the order parameter during the relaxation process \cite{fukuyama1971fluctuation,ussishkin2002gaussian,larkin2005theory}. The Langevin force $\zeta_{\mathbf{p}}$ is itself a vector in the internal pairing space, and is taken to be a Gaussian random force with zero mean $\langle \zeta_{\mathbf{p}}(t)\rangle=0$. The fluctuation--dissipation relation fixes its covariance \cite{wakatsuki2017nonreciprocal}
\begin{equation}
    \left\langle
    \zeta_{i,\mathbf{p}}(t)
    \zeta_{j,\mathbf{p}'}^\ast(t')
    \right\rangle
    =
    2\Xi'
    \delta_{ij}
    \delta_{\mathbf{p},\mathbf{p}'}
    \delta(t-t'),
\end{equation}
where $\left\langle\cdots\right\rangle$ denotes the ensemble average over the configurations of Langevin noise.

To better expose the geometric structure of the pairing fluctuation transport, here we introduce the non-equilibrium pairing fluctuation density matrix (PFDM) of $\Delta_{\mathbf{p}}(t)$ as \(\hat C_{\mathbf{p}}(t)  
    =
    \left\langle
    \Delta_{\mathbf{p}}(t)
    \Delta_{\mathbf{p}}^\dagger(t)
    \right\rangle\) \cite{mishonov2003kinetics}, which is parallel to the single-particle reduced density matrix for normal electron transport \cite{mandal2024quantum}. Thus the fluctuation-mediated current density can be defined as
\begin{equation}\label{eq:curr_def}
    J_a(t)
    =
    \frac{T}{V}\sum_{\mathbf{p}}
    \mathrm{tr}_{\mathrm{mat}}
    \left[
    \hat J_a(\mathbf{p}-e^*\mathbf{A}(t))\hat C_{\mathbf{p}}(t)
    \right].
\end{equation}
where $\hat J_a
    \bigl(\mathbf{p}\bigr)
    =e^\ast\partial_{p_a}\hat L_0^{-1}(\mathbf{p})$ is the current operator, \(\mathrm{tr}_{\mathrm{mat}}\) denotes the trace over internal pair-component indices only. As derived in \textbf{Supplementary Materials Note I}~\cite{SM}, following the multicomponent TDGL equation Eq.~\eqref{eq:tdgl_scalar_relaxation}, the kinetic equation obeyed by the PFDM is
\begin{align}
   \partial_t\hat C_{\mathbf{p}}(t)
    =
    &-
    \frac{1}{\Xi}
    \hat L_0^{-1}
    \bigl(\mathbf{p}-e^*\mathbf{A}(t)\bigr)
    \hat C_{\mathbf{p}}(t)\nonumber\\
    &-
    \frac{1}{\Xi^\ast}
    \hat C_{\mathbf{p}}(t)
    \hat L_0^{-1}
    \bigl(\mathbf{p}-e^*\mathbf{A}(t)\bigr)
    +
    \frac{2\Xi'}{|\Xi|^2}
    \hat 1.    
\label{eq:C_dynamics_final}
\end{align}
    Here, if we consider a static and uniform electric field, when $t\rightarrow\infty$ the system will reach a DC steady state. The DC steady state is not characterized by
\(\partial_t\hat C_{\mathbf{p}}=0\) at fixed canonical momentum. Rather, the steady-state PFDM is a function of the instantaneous kinetic
momentum: $\hat C_{\mathbf{p}}(t)=\hat C_{\mathbf{p}+e^\ast\mathbf{E} t}$.
Thus \(J_a
    =
    \frac{T}{V}\sum_{\mathbf p}
    \mathrm{tr}_{\mathrm{mat}}
    \left[
    \hat J_a(\mathbf p)\hat C_{\mathbf p}
    \right]\) is no longer time-dependent by relabeling \(\mathbf{p}+e^\ast\mathbf{E} t\rightarrow\mathbf{p}\) in Eq.~\eqref{eq:curr_def}, and the information of $\mathbf {E} $ is entirely encoded in $\hat C_{\mathbf{p}}$.

Using $\partial_t\hat C_{\mathbf{p}}(t)=e^\ast E_a\partial_{p_a}\hat C_{\mathbf{p}+e^\ast\mathbf{E} t}$, the steady-state kinetic equation of the PFDM $\hat C_{\mathbf{p}}$ becomes
\begin{equation}
    \mathscr{L}_\Xi [\hat{C}]
    =\frac{2\Xi'}{|\Xi|^2}\hat 1
    -e^\ast E_a\partial_a \hat{C}.
    \label{eq:steady_C_superoperator}
\end{equation}
where we define a superoperator for dissipative relaxation and precessional dynamics of operator $\hat X$: 
\begin{equation}
    \mathscr L_{\Xi}[\hat X]
    =
    \frac{\Xi'}{|\Xi|^2} \big\{ \hat X,\hat L_0^{-1} \big\}
    + i \frac{\Xi''}{|\Xi|^2} \big[\hat X,\hat L_0^{-1} \big] .
    \label{eq:LXi_def}
\end{equation}
 Eq.~\eqref{eq:steady_C_superoperator} states that field-induced drift in momentum space is balanced by TDGL relaxation and thermal noise. Then we expand the steady-state PFDM in powers of the electric field as
\(\hat C
    =
    \hat C^{(0)}
    +
    \hat C^{(1)}
    +\cdots\), with \(\hat C^{(n)}\sim E^n\).
The zeroth-order solution is $\hat C^{(0)}=\hat L_0$. We note that $\hat C^{(0)}$ plays the role of equilibrium density matrix \cite{ji2025density,guan2026exploring}. The relation $\hat C^{(0)}=\hat L_0$ implies that formulating the quantum information geometry of the pair propagator $\hat L_0$ is equivalent to formulating the one for $\hat C^{(0)}$. For \(n\geq 1\), the recursion relation is \(\mathscr L_\Xi[\hat C^{(n)}]
    =
    -
    e^\ast E_a\partial_a\hat C^{(n-1)}\).

    Similarly, we can  expand the response current to $J_a
    =J_a^{(1)}+ \cdots$,
in which \(J_a^{(n)}
    =
    \frac{T}{V}\sum_{\mathbf p}
    \mathrm{tr}_{\mathrm{mat}}
    \left[
    \hat J_a(\mathbf p)\hat C^{(n)}_{\mathbf p}
    \right]\) is the $n$-th order response current with $J_a^{(0)}=0$. To derive the concrete formula for \(J_a^{(n)}\), in the eigenmode basis, we have \(
    (\mathscr L_\Xi[\hat X])_{mn}
    =
    \mathcal K_{mn}
    X_{mn}\), in which \(\mathcal K_{mn}
    \equiv
\left(2\Xi'\bar\lambda_{mn}+i\Xi''\lambda_{mn}\right)/\left(|\Xi|^2\lambda_m\lambda_n\right)\). Then we can obtain the matrix element of the $n$-th order PFDM as
\begin{equation}
    \left(\hat C^{(n)}\right)_{mn}
    =\frac{-
    e^\ast E_a}{\mathcal K_{mn}}\left(\partial_a\hat C^{(n-1)}\right)_{mn},
\end{equation}
together with the matrix element of the current operator
\begin{equation}
    \left(\hat J_a\right)_{nm}
    =
    e^\ast
    \left(\partial_a\hat L^{-1}_0\right)_{nm}
    =
    -
    e^\ast
    \frac{\bar \lambda_{nm}(\hat{I}_a)_{nm}}{\lambda_n\lambda_m}.
    \label{eq:current_vertex_SLD}
\end{equation}
Finally, we arrive at the $n$-th order response current mediated by fluctuation as the result
\begin{equation}
 J_a^{(n)}
    =
    \frac{e^{\ast2}T}{V}E_b\sum_{mn,\mathbf p}
    \mathcal W_{mn}\left(\hat{I}_a\right)_{nm}\left(\partial_b\hat C^{(n-1)}\right)_{mn},
\label{eq:current}
\end{equation}
in which the weight matrix is defined as
\begin{equation}
    \mathcal W_{mn}
    =
    \frac{
    |\Xi|^2\bar\lambda_{mn}(
    2\Xi'\bar\lambda_{mn}-i\Xi''\lambda_{mn})
    }{
    (2\Xi'\bar\lambda_{mn})^2
    +(\Xi''\lambda_{mn})^2
    }.
    \label{eq:weight}
\end{equation}

\subsection{Geometric origin of paraconductivity}

Now we reveal that the paraconductivity of multicomponent superconductors has a quantum information geometric origin. For the linear response current \(J_a^{(1)}=\sigma_{ab}^{\text{para}}E_b\), combining Eq.~\eqref{eq:current} with Eq.~\eqref{eq:SLD}, we obtain
\begin{equation}
    \sigma^{\text{para}}_{ab}
    =
    \frac{e^{\ast2}T}{V}
    \sum_{mn,\mathbf p}
    \mathcal W_{mn}\bar \lambda_{mn}
    \left(\hat{I}_a\right)_{nm}
    \left(\hat{I}_b\right)_{mn}.
    \label{eq:sigma_linear_SLD_weight}
\end{equation}
We can show Eq.~\eqref{eq:sigma_linear_SLD_weight} is real, and the symmetric part is
\begin{align}
    \sigma^{\text{para}}_{(ab)}
    &=
    \frac{|\Xi|^2e^{\ast2}T}{2\Xi'V}
    \sum_{\mathbf p}
    \Bigg[
    \sum_n
    \frac{
    (\partial_a\lambda_n)(\partial_b\lambda_n)
    }{
    \lambda_n
    }
    \nonumber\\
    &+
    \sum_{m\neq n}
    \frac{
    (2\Xi')^2
    \bar{\lambda}_{mn}
    \lambda_{mn}^{2}
    }{
    (2\Xi'\bar{\lambda}_{mn})^2
    +
    (\Xi''\lambda_{mn})^2
    }
    g^{nm}_{ab}
    \Bigg]
    \label{eq:sigma_symmetric_metric_simplified}
\end{align}
whose intramode part recovers the result for single-component superconductors \cite{daido2024rectification}, and the quantum metric provides additional enhancement. 

The antisymmetric part of the paraconductivity is
\begin{equation}
    \sigma^{\text{para}}_{[ab]}=
    \frac{e^{\ast2}T}{2V}
    \sum_{mn,\mathbf p}
    \frac{
    |\Xi|^2\Xi''\lambda_{nm}^3
    }{
    \left(2\Xi'\bar\lambda_{nm}\right)^2
    +
    \left(\Xi''\lambda_{nm}\right)^2
    }
    \Omega^{nm}_{ab},
    \label{eq:sigma_antisymmetric_ordered_simplified}
\end{equation}
which gives rise to the fluctuational anomalous Hall effect, unique for multicomponent superconductors. 

Usually, a finite \(\Xi''\) is due to the particle-hole asymmetry near the Fermi energy in the normal bands \cite{fukuyama1971fluctuation,sumiyoshi2014giant,li2020fluctuational} and $|\Xi''|/ \Xi'\sim T_c/E_{\rm F}\ll1$ \cite{ebisawa1971wave}. Remarkably, within this realistic approximation regime, we establish a universal relation between the linear paraconductivity and the quantum information geometric tensors defined in Eq.~\eqref{eq:JL_spectral} and Eq.~\eqref{eq:MUC_spectral_Omega}, in which the symmetric part is
\begin{equation} 
     \sigma^{\text{para}}_{(ab)}=\frac{e^{\ast2}T\Xi'}{2V}
    \sum_{\mathbf p}
    \mathcal F_{ab},
    \label{eq:sigma_QFI}
\end{equation}
thus the longitudinal dissipative paraconductivity is controlled by the QFI matrix of the pairing fluctuations, while only the classical Fisher information contribution remains finite for single-component superconductors. The antisymmetric part is
\begin{equation} 
     \sigma^{\text{para}}_{[ab]}=\frac{e^{\ast2}T\Xi''}{V}
    \sum_{\mathbf p}
    \mathcal U_{ab},
    \label{eq:sigma_MUC}
\end{equation}
 which states the fluctuational anomalous Hall effect is controlled by the MUC \cite{leonforte2019uhlmann} and requires the reactive TDGL coefficient to be finite \cite{fukuyama1971fluctuation,ebisawa1971wave,sumiyoshi2014giant,li2020fluctuational}. Eq.~\eqref{eq:sigma_QFI} and Eq.~\eqref{eq:sigma_MUC} are the core results of this work. Therefore, we reveal an intrinsic quantum information geometric origin of the paraconductivity, especially for the fluctuational anomalous Hall effect, analogous to the Berry curvature origin of the normal state anomalous Hall effect \cite{xiao2010berry,nagaosa2010anomalous}.

\begin{figure}
		\centering
		\includegraphics[width=1\linewidth]{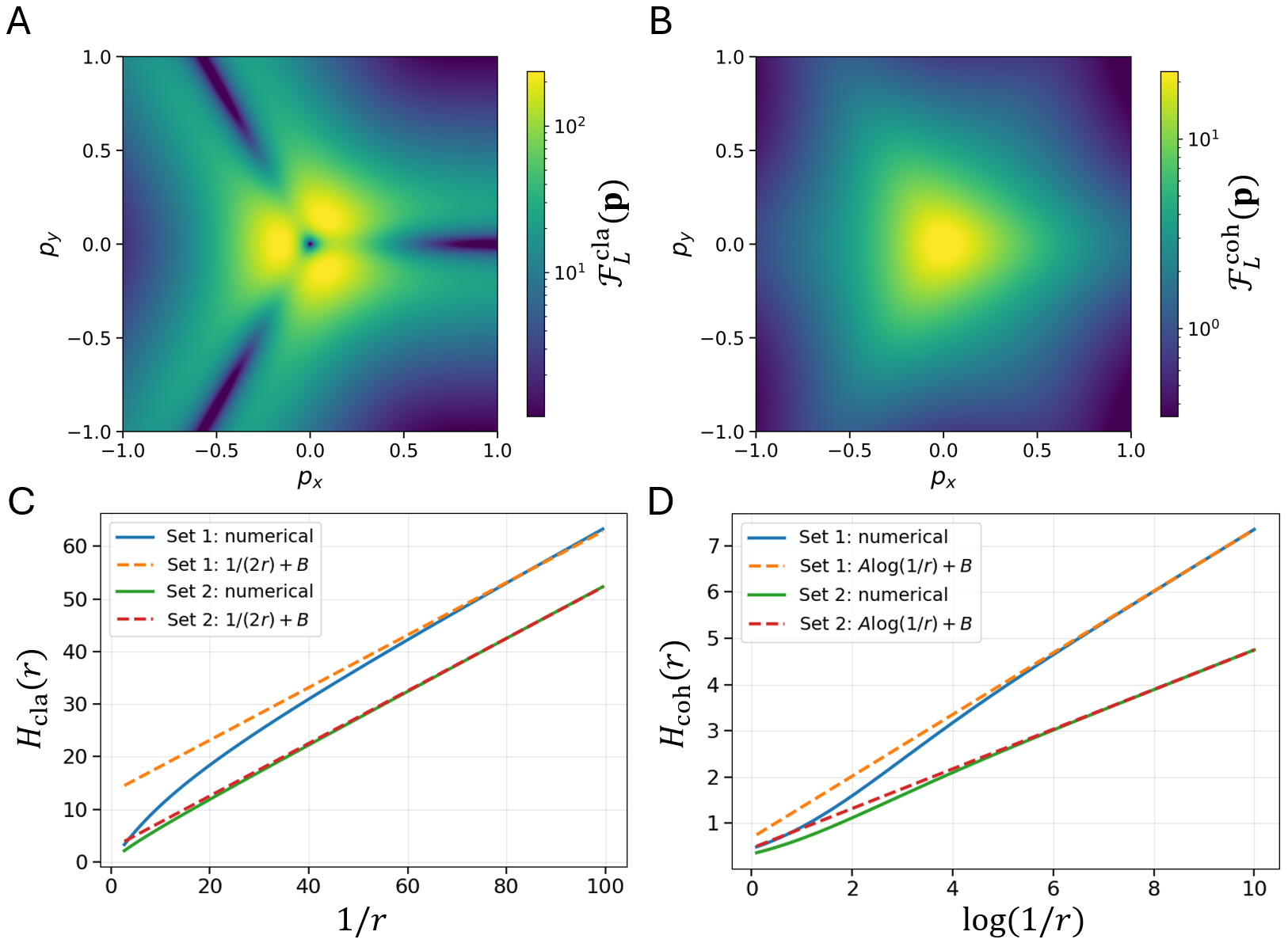}
		\caption{\textbf{Numerical results for longitudinal paraconductivity.} (A) and (B): Distribution of $\mathcal F^{\rm cla}_{L}(\mathbf p)$ and $\mathcal F^{\rm coh}_{L}(\mathbf p)$ for Eq.~\eqref{eq:R4G_minimal_kernel}. In both panels, the parameters are chosen as $\ell^2=0.8$, $g^2=0.5$, $\gamma_0=|\delta a|=1$, and $r=0.03$. (C) Dimensionless integral $H_{\rm cla}(r)$ as a function of $1/r$ for two parameter sets: (1) with $\ell^2=0.8$ and $g^2=0.5$, and (2) with $\ell^2=0.6$ and $g^2=0.4$. The dashed curves show the leading critical form $A(1/r)+B$, confirming the A-L divergence of the spectral contribution. (D) Dimensionless integral $H_{\rm coh}(r)$ as a function of $\log(1/r)$ for the same two parameter sets. The dashed curves show the leading form $A\log(1/r)+B$, demonstrating that the coherent contribution has only a logarithmic critical enhancement. $B$ denotes a non-universal contribution.
        } 
		\label{fig:fig2}
\end{figure}

Because $\sigma^{\text{para}}_{(ab)}$ and $\sigma^{\text{para}}_{[ab]}$ have a unified geometric origin, we can naturally obtain some bounds between them \cite{ozawa2021relations,onishi2024fundamental,shinada2025quantum}. In a two-dimensional superconductor, as proven in \textbf{Supplementary Materials Note II}~\cite{SM}, we have the inequality
\begin{equation}
4\left|
        n_U
    \right|\leq
    {\rm Tr}
    \sum_{\mathbf p}
    \mathcal F^{\rm coh}_{ab}(\mathbf p),
    \label{eq:integrated_geometric_hierarchy}    
\end{equation}
where $n_U=\sum_{\mathbf p}\mathcal U_{xy}(\mathbf p)$ is the Uhlmann number, as the
finite-temperature generalization of the Chern number in two-dimensional
fermionic systems~\cite{leonforte2019uhlmann}.

If the system has threefold or higher rotational symmetry, $\sigma^{\text{para}}_{L}=\sigma^{\text{para}}_{(xx)}=\sigma^{\text{para}}_{(yy)}$, $\sigma^{\text{para}}_{(xy)}=0$ and $\sigma^{\text{para}}_{H}=\sigma^{\text{para}}_{[xy]}$. We can separate two different contributions in $\sigma^{\text{para}}_{L}$ as $\sigma^{\text{para}}_{L}=\sigma^{\rm cla}_L+\sigma^{\rm coh}_L$, with $\sigma^{\text{cla}}_{L}\propto{\rm Tr}
    \sum_{\mathbf p}
    \mathcal F^{\rm cla}_{ab}$ and $\sigma^{\text{coh}}_{L}\propto{\rm Tr}
    \sum_{\mathbf p}
    \mathcal F^{\rm coh}_{ab}$. So $\sigma^{\text{para}}_{H}\propto n_U$ is bounded by $|\sigma^{\text{para}}_{H}/\Xi''|\leq\sigma^{\rm coh}_L/\Xi'\leq\sigma^{\rm para}_L/\Xi'$. This inequality can also be interpreted as an upper bound on the fluctuation anomalous Hall angle. We define two angles of fluctuational anomalous Hall effect \(\tan\theta_H^{\rm para}\equiv\sigma_H^{\rm para}/\sigma_L^{\rm para}\) and \(\tan\theta_H^{\rm coh}\equiv\sigma_H^{\rm para}/\sigma_L^{\rm coh}\). Then the inequality reads
\begin{equation}
    |\theta_H^{\rm para}|\leq|\theta_H^{\rm coh}|
    \leq
    \arctan\left(\frac{|\Xi''|}{\Xi'}\right)\approx\frac{|\Xi''|}{\Xi'}.
\end{equation}
Thus the anomalous Hall angle of paraconductivity is bounded by the ratio between the reactive and dissipative coefficients of the TDGL dynamics, as a result of quantum information geometry. This result has no direct counterpart in normal-state transport, e.g., in a Chern insulator with non-zero Chern number $C$, it can be easily violated for $\sigma_{L}=0$ while $\sigma_{H}= C e^2/h$ \cite{chang2023colloquium}.

\subsection{Application in the quarter-metal chiral superconductivity}
Recent experiments on rhombohedral tetralayer \cite{han2025signatures,sheekey2026visualizing} and pentalayer graphene \cite{dutta2026reconfigurable} reported signatures of chiral superconductivity. The superconducting states appear in gate-induced flat conduction bands and are developed in proximity to a spin-valley-polarized quarter-metal normal state. These observations motivate a minimal theory of chiral superconducting fluctuations in a single spin-valley-polarized conduction band, where time-reversal symmetry is already broken by the parent quarter-metal state \cite{geier2025chiral}.

We therefore describe the low-energy superconducting fluctuations by two chiral order-parameter components,
\(\boldsymbol{\Delta}
    =\left(\Delta_+,\Delta_-\right)\),
where $\Delta_+$ and $\Delta_-$ denote the two local chiral pairing channels with angular momentum $\pm1$ with the form factors
$f_{+}(\mathbf k)\sim e^{+i\theta_{\mathbf k}}$ and $f_{-}(\mathbf k)\sim e^{-i\theta_{\mathbf k}}$, respectively \cite{zhu2026microscopic}.
For the single simply connected Fermi surface region in the weak-pairing limit, the gap winding equals the BdG Chern number, thus these two channels can be identified with the two opposite BdG Chern sectors, $C=\pm1$ \cite{geier2025chiral,yoon2025quarter}.

The minimal continuum inverse propagator consistent with the chiral basis, the $C_3$ symmetry, and the spontaneous time-reversal symmetry breaking of rhombohedral graphene is
\begin{equation}
     \hat L_0^{-1}(\mathbf p)
    =
    \begin{pmatrix}
        a_+ + \gamma_+ p^2
        &
        \gamma_m p_+^2+\lambda p_-
        \\
        \gamma_m^* p_-^2+\lambda^* p_+
        &
        a_- + \gamma_- p^2
    \end{pmatrix},
\label{eq:R4G_minimal_kernel}
\end{equation}
where $p_{\pm}=p_x\pm ip_{y}$. Each term in Eq.~\eqref{eq:R4G_minimal_kernel} has a direct physical origin. First, $a_+\neq a_-$ and $\gamma_+\neq \gamma_-$ represent the fact that the two chiral channels are no longer degenerate in a valley-polarized, time-reversal-breaking normal state. Microscopically, this splitting originates from the quantum geometry of the host conduction band together with the pairing interaction projected onto that low-energy band \cite{zhu2026microscopic}. Second, the off-diagonal term $\gamma_m p_+^2$ is the leading rotationally invariant mixing between the two chiral components, and $|\gamma_m|<\gamma_0$. The linear-gradient term $\lambda p_-$ is a Lifshitz invariant \cite{nagashima2026type} allowed by the crystalline $C_{3\perp}$ symmetry but requires time-reversal symmetry breaking in rhombohedral graphene. For the convenience of later discussions, we parametrize $\delta a=(a_{+}-a_{-})/2$, $\bar a=(a_{+}+a_{-})/2$, and we assume $\gamma_{0}=\gamma_{+}=\gamma_{-}$. We further rescale the parameters as \(g\equiv|\gamma_m|/\gamma_0\), \(\ell^2\equiv |\lambda|^2/(\gamma_0|\delta a|)\). To expose the critical temperature dependence, we introduce the dimensionless distance from the superconducting transition \(r\equiv (\bar a-|\delta a|)/|\delta a|\propto (T-T_c)/T_c>0\), so that the transition occurs at $r=0$.

\begin{figure}
		\centering
		\includegraphics[width=1\linewidth]{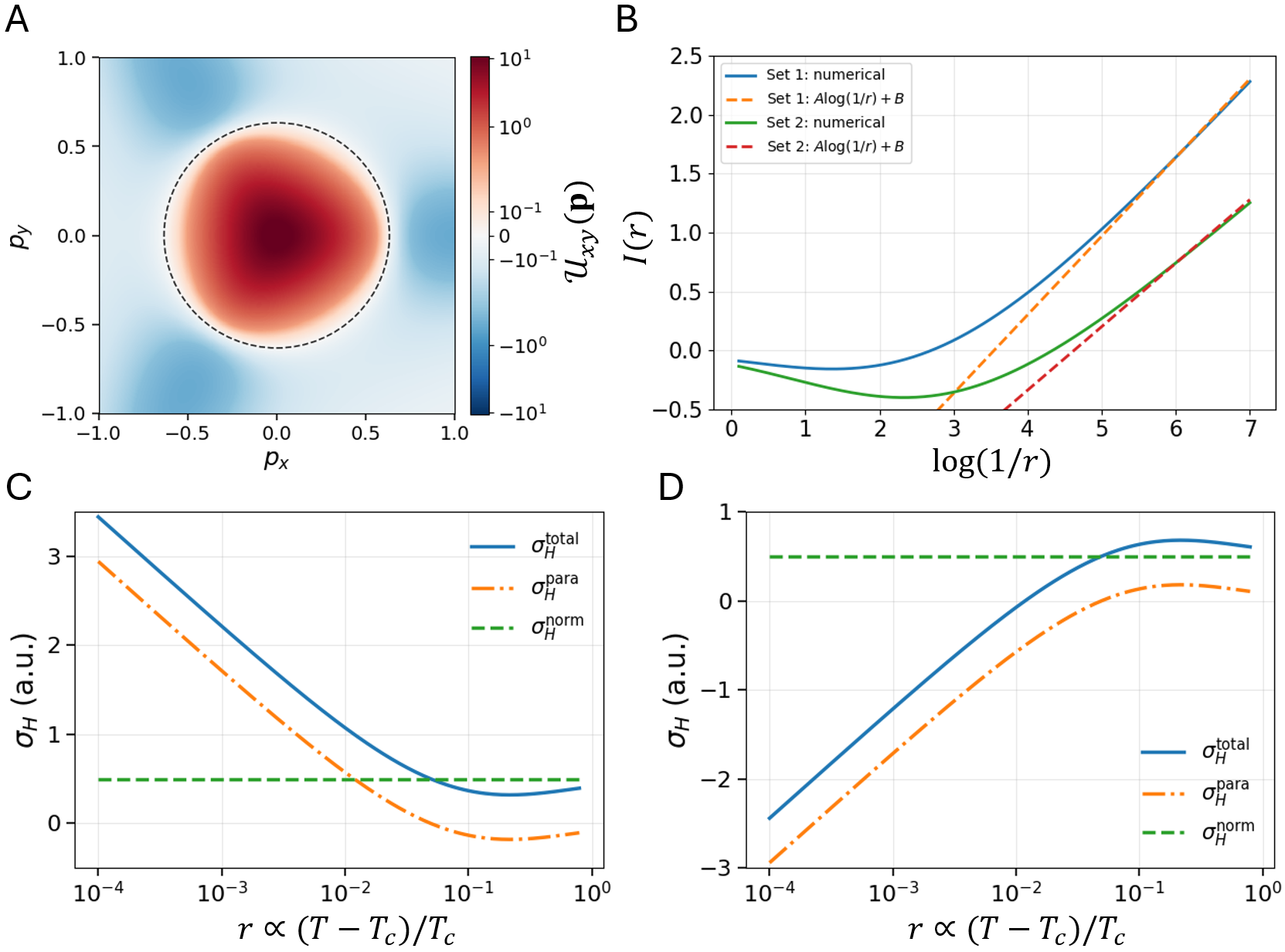}
		\caption{\textbf{Results for fluctuational anomalous Hall effects.} (A) Momentum-resolved MUC $\mathcal U_{xy}(\mathbf p)$ for Eq.~\eqref{eq:R4G_minimal_kernel}. The parameters are chosen as $\ell^2=0.8$, $g^2=0.5$, $\gamma_0=|\delta a|=1$, and $r=0.03$. The dashed circle marks the zero of the numerator, across which $\mathcal U_{xy}$ changes sign. (B) Dimensionless integral $I(r)$ as a function of $\log(1/r)$ for two parameter sets: (1) $\ell^2=0.8$ and $g^2=0.5$, (2) $\ell^2=0.7$ and $g^2=0.6$. The agreement at large $\log(1/r)$ confirms the critical behavior of $I(r)$. Panel (C) corresponds to a local-attraction scenario with a positive $\sigma_H^{\rm norm}$ background, while panel (D) corresponds to a Coulomb-repulsion scenario. In both cases, the fluctuational anomalous Hall conductivity $\sigma_H^{\rm para}$ grows logarithmically upon approaching $T_c$, and its competition with $\sigma_H^{\rm norm}$ can strongly reshape the observed anomalous Hall response $\sigma_H^{\rm tot}$. In (C), it can be enhanced, whereas in (D) it may decrease and even change sign upon approaching the superconducting transition.}
		\label{fig:fig3}
\end{figure}

We first discuss the longitudinal paraconductivity. Because the system possesses $C_{3\perp}$ symmetry, \(\sigma^{\text{para}}_{L}\propto\sum_{\mathbf p}(\mathcal F_{xx}+\mathcal F_{yy})/2\). We can define $\mathcal F_{L}=(\mathcal F_{xx}+\mathcal F_{yy})/2$ and plot the distributions of two main components $\mathcal F^{\rm cla}_{L}(\mathbf p)$ and $\mathcal F^{\rm coh}_{L}(\mathbf p)$ from Eq.~\eqref{eq:JL_spectral} separately in Fig.~\ref{fig:fig2}~(A) and (B). We rewrite the two corresponding longitudinal paraconductivities as
\begin{equation}
    \sigma_L^{\rm cla/\rm coh}
    =
    \frac{e^{*2}T\Xi'}{4\pi|\delta a|}
    H_{\rm cla/\rm coh}(r),
    \label{eq:sigma_L_cla_coh_dimensionless}
\end{equation}
where \(H_{\rm cla}(r)\) and \(H_{\rm coh}(r)\) are dimensionless
functions of \(r\). The details can be found in \textbf{Supplementary Materials Note III}~\cite{SM}. Near the critical temperature $r\rightarrow0^+$, the classical sector becomes
\begin{align}
    H_{\rm cla}(r)
\simeq
\frac{\kappa_{\ell}^2}{\pi}
\int &d^2\mathbf k
\frac{k^2}
{\left(r+\kappa_{\ell}k^2\right)^3}
\rightarrow
\frac{1}{2r},\nonumber\\\text{and}\quad\sigma_L^{\rm cla}
    &\rightarrow 
     \frac{e^{*2}T_c\Xi'}{8\pi|\delta a|r},
    \label{eq:J_cla_divergence}
\end{align} 
 with \(\mathbf k=\sqrt{\gamma_0/|\delta a|}\mathbf p\) and $\kappa_{\ell}=1-\ell^2/2$. For the coherent sector,
\begin{align}
H_{\rm coh}(r)
    \simeq
    &\frac{\ell^2}{2\pi}\int \frac{d^2\mathbf k}{r+\kappa_\ell k^2}
    \rightarrow
    \frac{\ell^2}{2\kappa_\ell}\log\frac{1}{r}+\mathcal O(1),\nonumber\\
    \text{and}&\quad\sigma_L^{\rm coh}\rightarrow \frac{
        e^{*2}T_c\ell^2\Xi'
    }{
        8\pi |\delta a|\kappa_\ell}\log\frac{1}{r}.    
    \label{eq:J_cla_divergence}
\end{align}
The agreement between exact $H_{\rm cla/\rm coh}(r)$ and their critical values is checked numerically in Fig.~\ref{fig:fig2}~(C) and (D), respectively. Thus the two longitudinal channels have distinct critical singularities. The leading \(1/r\) divergence of the total longitudinal paraconductivity is therefore carried by the classical spectral sector, which is consistent with previous studies \cite{larkin2005theory,lundemo2026fluctuation}. The coherent sector exhibits only a logarithmic enhancement. Since $ \ell^2\propto |\lambda|^2
$, this divergence is relevant to the Lifshitz invariant rather than the quadratic coupling between two chiral branches.

Then we are interested in the intrinsic fluctuational anomalous Hall effect in Eq.~\eqref{eq:R4G_minimal_kernel}. Using Eq.~\eqref{eq:MUC_spectral_Omega}, the distribution of $\mathcal U_{xy}(\mathbf p)$ is presented in Fig.~\ref{fig:fig3}~(A). The fluctuational anomalous Hall conductivity can be rewritten as
\begin{equation}
    \sigma^{\rm para}_{H}
    =
    \frac{e^{*2}T\Xi''}{4\pi}
    \frac{\delta a}{|\delta a|^2}
     I(r),
     \label{eq:integral_para}
\end{equation}
where $I(r)$ is a dimensionless function of $r$, as shown in \textbf{Supplementary Materials Note III}~\cite{SM}. When $r\rightarrow0^+$, the leading singularity follows 
\begin{equation}
    I(r)
    \rightarrow
    \frac{\ell^2}{2-\ell^2}
    \log\frac{1}{r}+\mathcal O(1),
    \label{eq:dimensionless_diverg}
\end{equation}
with the same infrared origin as $H_{\rm coh}(r)$. Microscopic calculation \cite{qin2026chiral} suggests that the Lifshitz invariant is not large enough to induce an incommensurate finite-momentum pairing instability, i.e., \(g^2<\kappa_{\ell}\). The agreement between exact $I(r)$ and Eq.~\eqref{eq:dimensionless_diverg} is checked numerically in the large-\(\ln(1/r)\) regime, as shown in Fig.~\ref{fig:fig3}~(B).

Equivalently, near $T_c$, the fluctuational anomalous Hall conductivity becomes
\begin{equation}
    \sigma_{H}^{\rm \text{para}}
    \approx
    \frac{
        e^{*2}T_c\Xi''\delta a\ell^2
    }{
        8\pi |\delta a|^2\kappa_\ell} \ln\frac{T_c}{T-T_c}+B,
    \label{eq:hall_para_lifshitz_diverg}
\end{equation}
where $B$ is a non-universal contribution. Eq.~\eqref{eq:hall_para_lifshitz_diverg} shows that the fluctuational anomalous Hall conductivity is also logarithmically divergent with respect to $T-T_c$ \cite{efimkin2021topological}. Since this logarithm is a marginal two-dimensional singularity, we do not expect the same divergence in a truly three-dimensional bulk system. More interestingly, if we only look at the leading divergent part of $\sigma_{H}^{\rm \text{para}}$ and $\sigma_{L}^{\rm \text{coh}}$, we find that $|\sigma^{\text{para}}_{H}/\Xi''|\approx\sigma^{\rm coh}_L/\Xi'$ and $|\theta_H^{\rm coh}|\approx|\Xi''|/\Xi'$, i.e., the geometric bound becomes equality near the critical point. This fact reveals the unified coherent multicomponent origin of both $\sigma_{H}^{\rm \text{para}}$ and $\sigma_{L}^{\rm \text{coh}}$.

\section{Discussion}

We now discuss the experimental implications of the fluctuational anomalous Hall response Eq.~\eqref{eq:hall_para_lifshitz_diverg} in rhombohedral graphene. The sign of the diverging $\sigma_{H}^{\rm \text{para}}$ is determined by $\text{sgn}(\sigma_{H}^{\rm \text{para}})=\text{sgn}(\Xi'')\text{sgn}(\delta a)$. First, overscreened repulsive interactions of the Kohn-Luttinger type tend to favor a BdG Chern number with a sign opposite to that of the averaged Berry curvature $\bar{\mathcal{B}}$ of the parent conduction band, while the short-range attraction favors the same sign \cite{may2026pairing}. Phenomenologically, this microscopic information is encoded in $\text{sgn}(\delta a)=\pm\text{sgn}(\bar{\mathcal{B}})$, where the $+$ ($-$) sign corresponds to the repulsive (attractive) interaction, respectively. Second, in the weak coupling limit, the sign of the reactive TDGL coefficient is related to the dependence of the transition temperature on the chemical potential, \({\rm sgn}\,\Xi''
    =
    -\,{\rm sgn}
    \left(
        \partial T_c/\partial\mu
    \right)\) \cite{aronov1995gauge,michaeli2012hall}.
When rhombohedral graphene is doped away from the band edge, $T_c$ increases with carrier density as the system approaches the superconducting dome, thus $\Xi''<0$ in this regime. Conversely, $T_c$ decreases as it moves further away from the dome, so one expects $\Xi''>0$ in this case \cite{han2025signatures,geier2025chiral}.

However, the background anomalous Hall conductivity in the normal state is $\sigma_{H}^{\rm norm}\propto-\bar{\mathcal{B}}$. As a result, fluctuational anomalous Hall conductivity can have the same sign as, or the opposite sign to, the normal anomalous Hall background $\sigma_{H}^{\rm \text{norm}}$ when $T$ approaches $T_c$, depending on the pairing mechanism. In this sense, although $\sigma_{H}^{\rm \text{para}}$ is weakly divergent, it can still play an important role in the total anomalous Hall conductivity $\sigma_{H}^{\rm \text{tot}}=\sigma_{H}^{\rm \text{norm}}+\sigma_{H}^{\rm \text{para}}$ near the superconducting transition and provides a diagnostic for both the chiral pairing nature and the pairing mechanism. As a representative scenario, here we assume the experiment is performed in the single simply connected Fermi surface region with a finite BdG Chern number and away from the superconducting dome, for which $\Xi''>0$. In this scenario, for local attractive interaction, $\sigma_{H}^{\rm \text{para}}$ has the same sign as $\sigma_{H}^{\rm \text{norm}}$, so $\sigma_{H}^{\rm \text{tot}}$ can be enhanced approaching the critical temperature, as shown in Fig.~\ref{fig:fig3}~(C). For overscreened Coulomb repulsion, they have opposite signs, so
\(\sigma_H^{\rm para}\) may reduce the magnitude of
\(\sigma_H^{\rm tot}\), or even invert its sign as the temperature decreases toward the superconducting transition, as shown in Fig.~\ref{fig:fig3}~(D). We note that, to make a more detailed comparison with the experimental results, one needs a more careful investigation of the Fermiology of the parent state \cite{kalantre2026fermiology}.

Several open questions are left for future investigation. Although nonlinear and nonreciprocal paraconductivity in single-component superconductors has been extensively studied \cite{wakatsuki2017nonreciprocal,daido2024rectification,matsumoto2025reciprocal,dong2025enhanced}, a systematic quantum information geometric interpretation for multicomponent superconductors remains unexplored. We envision that a similar geometric framework can be established for second-order nonlinear paraconductivity, in analogy to the Berry curvature dipole \cite{sodemann2015quantum,zhang2023higher} and Berry connection polarizability \cite{gao2014field,wang2021intrinsic,liu2021intrinsic} interpretations of normal-state nonlinear Hall effects. Notably, the nonlinear paraconductivity can exhibit stronger singularities than its linear counterpart, rendering it highly accessible to experimental verification. Furthermore, thermal transport \cite{ussishkin2002gaussian} and thermoelectric effects \cite{sumiyoshi2014giant,li2020fluctuational} mediated by multicomponent superconducting fluctuations merit further exploration to uncover their underlying quantum information geometry.

Possible future directions include extending the present paradigm to other many-body fluctuations beyond superconducting fluctuations, such as excitonic superfluidity \cite{efimkin2021topological}, magnonic \cite{matsumoto2011rotational} or phononic \cite{qin2012berry} thermal Hall systems, density-wave \cite{pokharel2022dynamics} or nematic fluctuations \cite{chu2012divergent}. Finally, in this work, we did not consider the possible normal-state band quantum geometric contribution to the paraconductivity, which may also play a role in multi-band superconductors \cite{chen2025generalized}.

In summary, our results identify superconducting fluctuation transport as a direct probe of quantum information geometry beyond single-particle Bloch states. The longitudinal paraconductivity measures the QFI of the pairing-fluctuation manifold, while the fluctuational anomalous Hall effect measures its MUC. This establishes a route to detect mixed-state geometry in collective many-body matter through transport.

\section*{Materials and Methods}

We used the multicomponent TDGL approach in the Gaussian fluctuation regime. The derivation of the PFDM kinetic equation, the quantum information geometric bounds, and the critical asymptotics of the paraconductivity are given in the \textbf{Supplementary Materials}~\cite{SM}.

\section*{Acknowledgments}
We thank Koki Shinada for inspiring discussions.

\subsection*{Funding} 

N.N. was supported by JSPS KAKENHI Grant Numbers 24H00197, 24H02231 and 24K00583. N.N. was supported by the RIKEN TRIP initiative. Y.M.X. acknowledges financial support from the RIKEN Special Postdoctoral
Researcher (SPDR) Program. 

\subsection*{Author contributions}  

Conceptualization: Z.T.S., Y.M.X. and N.N. Methodology: Z.T.S. Investigation: Z.T.S. Writing–original draft: Z.T.S. Writing–review and editing: Z.T.S., Y.M.X. and N.N. Funding acquisition: Y.M.X. and N.N.

\subsection*{Competing interests} 

The authors declare that they have no competing interests.

\subsection*{Data, code and materials availability}

All data needed to evaluate and reproduce the results in the paper are present in the paper and/or the Supplementary Materials. This study did not generate new materials.

\clearpage
		\onecolumngrid
\begin{center}
		\textbf{\large Supplementary Materials for ``Quantum Information Geometry of Multicomponent Superconducting Fluctuation Transport''}\\[.2cm]
		 Zi-Ting Sun,$^{1}$  Ying-Ming Xie,$^{1}$  Naoto Nagaosa$^{1,2}$\\[.1cm]
		
        {\itshape ${}^1$RIKEN Center for Emergent Matter Science (CEMS), Wako, Saitama 351-0198, Japan}
        
        {\itshape ${}^2$Fundamental Quantum Science Program (FQSP), TRIP Headquarters, RIKEN, Wako 351-0198, Japan}

\end{center}

\setcounter{equation}{0}
\setcounter{section}{0}
\setcounter{figure}{0}
\setcounter{table}{0}
\setcounter{page}{1}
\renewcommand{\theequation}{S\arabic{equation}}
\renewcommand{\thesection}{ \Roman{section}}

\renewcommand{\thefigure}{S\arabic{figure}}
\renewcommand{\thetable}{\arabic{table}}
\renewcommand{\tablename}{Supplementary Table}

\renewcommand{\bibnumfmt}[1]{[S#1]}
\makeatletter

\maketitle
\section*{\bf{\uppercase\expandafter{Supplementary Note I: Derivation of the kinetic equation for the pairing fluctuation density matrix}}}

In this section, we show how to derive the kinetic equation for the PFDM. First, we describe how to determine the multicomponent TDGL equation  \cite{schmid1969diamagnetic}. We denote $\hat{\mathcal{L}}^{-1}(i\Omega_n,\mathbf p;\mathbf A)$ as the inverse propagator in the external electromagnetic field. The real-time dynamics is governed by the retarded inverse pair propagator, obtained by analytic continuation, \(\hat{\mathcal{L}}_R^{-1}(\omega,\mathbf p;\mathbf A)
    =
    \hat{\mathcal{L}}^{-1}(i\Omega_n,\mathbf p;\mathbf A)
    \big|_{i\Omega_n\rightarrow \omega+i0^+}\).
In the low-frequency and long-wavelength approximation, we can expand it as
\begin{equation}
\hat{\mathcal{L}}_R^{-1}(\omega,\mathbf p;\mathbf A)
    \approx
    \hat L_0^{-1}(\mathbf p-e^*\mathbf A)
    -
    i\omega\Xi\hat 1   
\end{equation}
In a general multicomponent problem, the kinetic coefficient can be a
matrix \(\hat\Xi\).  In the \textbf{Main Text} and below, we adopt the minimal
relaxation structure \(
    \hat\Xi=\Xi\hat 1,\) where \(\Xi=\Xi'+i\Xi''\) is a scalar complex coefficient. Fourier transforming $\hat{\mathcal{L}}_R^{-1}(\omega,\mathbf p;\mathbf A)$ in this approximation back to real time, and considering the Langevin noise, we obtain the multicomponent TDGL equation:
\begin{equation}
    \Xi \partial_t \Delta_{\mathbf{p}}(t)
    =
    -
    \hat L_0^{-1}
    \bigl(\mathbf{p}-e^*\mathbf{A}(t)\bigr)
    \Delta_{\mathbf{p}}(t)
    +
    \zeta_{\mathbf{p}}(t).
    \label{eq:tdgl_scalar_relaxation}
\end{equation}

For a spatially uniform DC electric field, we define
\begin{equation}
    \hat H_{\mathbf p}(t)
    \equiv
    \hat L_0^{-1}
    \bigl(\mathbf p+e^\ast\mathbf E t\bigr).
\end{equation}
The multicomponent TDGL equation then becomes
\begin{equation}
    \partial_t\Delta_{\mathbf p}(t)
    =
    -
    \frac{1}{\Xi}
    \hat H_{\mathbf p}(t)
    \Delta_{\mathbf p}(t)
    +
    \eta_{\mathbf p}(t),
    \quad
    \eta_{\mathbf p}(t)
    \equiv
    \frac{1}{\Xi}\zeta_{\mathbf p}(t).
    \label{eq:tdgl_eta_complex_xi}
\end{equation}

We now derive the kinetic equation obeyed by the PFDM of the multicomponent fluctuating order parameter. To handle the white noise carefully, we write the stochastic equation
over an infinitesimal time interval \(dt\):
\begin{equation}
    d\Delta_{\mathbf p}
    =
    -
    \frac{1}{\Xi}
    \hat H_{\mathbf p}(t)
    \Delta_{\mathbf p}(t)\,dt
    +
    dW_{\mathbf p}(t),
    \label{eq:dDelta_ito}
\end{equation}
where the Wiener increment is defined as
\begin{equation}
    dW_{\mathbf p}(t)
    \equiv
    \int_t^{t+dt}ds\,\eta_{\mathbf p}(s),
\end{equation}
which satisfies
\begin{equation}
    \left\langle
    dW_{\mathbf p}(t)
    dW_{\mathbf p'}^\dagger(t)
    \right\rangle
    =
    \frac{2\Xi'}{|\Xi|^2}
    \delta_{\mathbf p,\mathbf p'}
    \hat 1\,dt .
    \label{eq:dW_correlation}
\end{equation}
In the Ito convention, the increment \(dW_{\mathbf p}(t)\) is
uncorrelated with the order parameter at the beginning of the time step:
\begin{equation}
    \left\langle
    dW_{\mathbf p}(t)
    \Delta_{\mathbf p}^\dagger(t)
    \right\rangle
    =
    0,
    \qquad
    \left\langle
    \Delta_{\mathbf p}(t)
    dW_{\mathbf p}^\dagger(t)
    \right\rangle
    =
    0 .
    \label{eq:ito_independence}
\end{equation}
Then we have
\begin{align}
    &\hat C_{\mathbf p}(t+dt)
    =
    \left\langle
    \left(
    \Delta_{\mathbf p}+d\Delta_{\mathbf p}
    \right)
    \left(
    \Delta_{\mathbf p}+d\Delta_{\mathbf p}
    \right)^\dagger
    \right\rangle
    \nonumber\\
    &=
    \hat C_{\mathbf p}(t)
    +
    \left\langle
    d\Delta_{\mathbf p}
    \Delta_{\mathbf p}^\dagger
    \right\rangle
    +
    \left\langle
    \Delta_{\mathbf p}
    d\Delta_{\mathbf p}^\dagger
    \right\rangle
    +
    \left\langle
    d\Delta_{\mathbf p}
    d\Delta_{\mathbf p}^\dagger
    \right\rangle .
    \label{eq:C_increment}
\end{align}
The first two terms give the deterministic relaxation:
\begin{align}
    \left\langle
    d\Delta_{\mathbf p}
    \Delta_{\mathbf p}^\dagger
    \right\rangle
    &=
    -
    \frac{1}{\Xi}
    \hat H_{\mathbf p}(t)
    \hat C_{\mathbf p}(t)\,dt,
    \\
    \left\langle
    \Delta_{\mathbf p}
    d\Delta_{\mathbf p}^\dagger
    \right\rangle
    &=
    -
    \frac{1}{\Xi^\ast}
    \hat C_{\mathbf p}(t)
    \hat H_{\mathbf p}(t)\,dt .
\end{align}
The last term is the Ito correction term:
\begin{equation}
    \left\langle
    d\Delta_{\mathbf p}
    d\Delta_{\mathbf p}^\dagger
    \right\rangle
    =
    \left\langle
    dW_{\mathbf p}
    dW_{\mathbf p}^\dagger
    \right\rangle
    =
    \frac{2\Xi'}{|\Xi|^2}
    \hat 1\,dt .
    \label{eq:ito_noise_correction}
\end{equation}
This term is essential.  It is the contribution from the white noise correlator at equal times and cannot be obtained by treating the noise as an ordinary differentiable function. 

Combining these terms and dividing by \(dt\), we obtain the kinetic equation of the PFDM:
\begin{align}
   \partial_t\hat C_{\mathbf{p}}(t)
    =
    &-
    \frac{1}{\Xi}
    \hat L_0^{-1}
    \bigl(\mathbf{p}-e^*\mathbf{A}(t)\bigr)
    \hat C_{\mathbf{p}}(t)\nonumber\\
    &-
    \frac{1}{\Xi^\ast}
    \hat C_{\mathbf{p}}(t)
    \hat L_0^{-1}
    \bigl(\mathbf{p}-e^*\mathbf{A}(t)\bigr)
    +
    \frac{2\Xi'}{|\Xi|^2}
    \hat 1.    
    \label{eq:C_dynamics_final}
\end{align}

Another way to obtain $\hat C_{\mathbf p}(t)$ is to directly solve the TDGL equation and substitute the solution into the definition of $\hat C_{\mathbf p}(t)$ \cite{wakatsuki2017nonreciprocal}. For a spatially uniform DC electric field, the formal steady-state solution $\Delta_{\mathbf p}(t)$ is
\begin{equation}
    \Delta_{\mathbf p}(t)
    =
    \int_{-\infty}^{t}
    \frac{dt'}{\Xi}
    \mathcal T
    \exp\left[
    -
    \frac{1}{\Xi}
    \int_{t'}^{t}
    d\tau\,
    \hat L_0^{-1}
    \bigl(\mathbf p+e^\ast\mathbf E \tau\bigr)
    \right]
    \zeta_{\mathbf p}(t'),
    \label{eq:tdgl_formal_solution}
\end{equation}
where $\mathcal T$ denotes time ordering. This expression describes
fluctuating Cooper pairs generated by thermal noise, accelerated by the
electric field during their finite lifetime, and relaxed by the TDGL damping.

\section*{\bf{\uppercase\expandafter{Supplementary Note II: Quantum information geometric bounds on paraconductivity}}}
 We discuss some quantum information geometric bounds on paraconductivity in this section. In two spatial dimensions, \(g^{nm}_{xx}g^{nm}_{yy}-\left(g^{nm}_{xy}\right)^2
    =\left(\Omega^{nm}_{xy}/2\right)^2\). Therefore
\begin{equation}
    2|\mathcal U^{nm}_{xy}|
    =
    \frac{
        |\lambda_{nm}|
    }{
        2\bar\lambda_{mn}
    }
    \sqrt{
        \det\mathcal F^{nm}_{ab}
    },
\end{equation}
where we define
\begin{equation}
    \mathcal F^{nm}_{ab}
    =
    \frac{
        \lambda_{nm}^{2}
    }{
        \bar\lambda_{mn}
    }
    g^{nm}_{ab},
    \qquad
    \mathcal U^{nm}_{ab}
    =
    \frac{
        \lambda_{nm}^{3}
    }{
        \bar\lambda_{mn}^{2}
    }
    \frac{
        \Omega^{nm}_{ab}
    }{
        8
    },
\end{equation}
and then \(\mathcal F^{\rm coh}_{ab}
    =
    \sum_{n\neq m}
    \mathcal F^{nm}_{ab}\) and \(\mathcal U_{ab}
    =
    \sum_{n\neq m}
    \mathcal U^{nm}_{ab}\).

Since \( |\lambda_{nm}|/
        \bar\lambda_{mn}\leq2\), each intermode contribution obeys \(2|\mathcal U^{nm}_{xy}|
    \leq
    \sqrt{
        \det\mathcal F^{nm}_{ab}
    }\). Summing over \(n\neq m\), we have \(2|\mathcal U_{xy}|
    \leq
    \sum_{n\neq m}
    2|\mathcal U^{nm}_{xy}|\) and \(\sum_{n\neq m}
    \sqrt{
        \det\mathcal F^{nm}_{ab}
    }\leq
    \sqrt{
        \det\mathcal F^{\rm coh}_{ab}
    }\). Here we have used the Minkowski determinant inequality for
two-dimensional positive semidefinite matrices, \(
    \sum_i \sqrt{\det A_i}
    \le
    \sqrt{\det\left(\sum_i A_i\right)} .
\)
Thus we obtain \(2|\mathcal U_{xy}|
    \leq\sqrt{
        \det\mathcal F^{\rm coh}_{ab}
    }\).

After summing over momentum, we can write the hierarchical inequalities
\begin{equation}
2\left|
        n_U
    \right|
    \leq
    2\sum_{\mathbf p}
    |\mathcal U_{xy}(\mathbf p)|\leq\mathcal V^{\rm coh}_F,
    \label{eq:integrated_geometric_hierarchy}    
\end{equation}
where
$n_U=\sum_{\mathbf p}\mathcal U_{xy}(\mathbf p)$ is the Uhlmann number, as the
finite-temperature generalization of the Chern number in two-dimensional
fermionic systems~\cite{leonforte2019uhlmann}. For two-dimensional superconductors, we note that
$\sigma^{\text{para}}_{H}\propto n_U$. We can also define the quantum volume of the coherent Fisher information \cite{ozawa2021relations} as $\mathcal V^{\rm coh}_F=\sum_{\mathbf p}
    \sqrt{
        \det\mathcal F^{\rm coh}_{ab}(\mathbf p)
    }$. Thus the fluctuational anomalous Hall conductivity gives a lower bound on $\mathcal V^{\rm coh}_F$,
\begin{equation}
    \mathcal V^{\rm coh}_F
    \geq
    \frac{2V}{e^{*2}T|\Xi''|}
    \left|
    \sigma^{\text{para}}_{[xy]}
    \right|.
    \label{eq:volume_lower_bound_hall}
\end{equation}
A finite fluctuational anomalous Hall effect, therefore, witnesses a nonzero
quantum volume of the pairing fluctuations.

Furthermore, 
\begin{equation}
    \mathcal V^{\rm coh}_F
    \leq
    \sqrt{
        \det
        \sum_{\mathbf p}
        \mathcal F^{\rm coh}_{ab}(\mathbf p)
    } \leq
    \frac12
    {\rm Tr}
    \sum_{\mathbf p}
    \mathcal F^{\rm coh}_{ab}(\mathbf p).
    \label{eq:integrated_geometric_hierarchy}
\end{equation}
If the system has threefold or higher rotational symmetry, $\sigma^{\text{coh}}_{L}\propto{\rm Tr}
    \sum_{\mathbf p}
    \mathcal F^{\rm coh}_{ab}$ so that
\begin{equation}
    \mathcal V^{\rm coh}_F\leq
    \frac{2V}{e^{*2}T\Xi'}
    \sigma^{\text{coh}}_{L}.
    \label{eq:volume_upper_bound_logi}
\end{equation}
Combining Eq.~\eqref{eq:volume_lower_bound_hall} and Eq.~\eqref{eq:volume_upper_bound_logi}, we can arrive at the geometric bounds presented in the \textbf{Main Text}. This hierarchy shows that the Uhlmann curvature is controlled first by
the coherent Fisher information, not by the full Fisher information. Consequently, the full Fisher
information tensor gives a looser bound because ${\rm Tr}
    \sum_{\mathbf p}
    \mathcal F^{\rm coh}_{ab}(\mathbf p)\leq{\rm Tr}
    \sum_{\mathbf p}
    \mathcal F_{ab}(\mathbf p)$ and $\sigma^{\text{coh}}_{L}\leq\sigma^{\text{para}}_{L}$.

As an example, for two-component superconductors, the inverse pair propagator is a two-by-two Hermitian matrix,
\begin{equation}
    \hat L_0^{-1}(\mathbf p)
    =
    d_0(\mathbf p)\sigma_0
    +
    \mathbf d(\mathbf p)\cdot\boldsymbol\sigma.
\end{equation}
The MUC reduces to
\begin{equation}
    \mathcal U_{xy}(\mathbf p)
    =
    \frac{
        |\mathbf d|^3
    }{
        d_0^2
        \left(
            d_0^2-|\mathbf d|^2
        \right)
    }
    \,
    \hat{\mathbf d}\cdot
    \left(
        \partial_{p_x}\hat{\mathbf d}
        \times
        \partial_{p_y}\hat{\mathbf d}
    \right).
    \label{eq:uxy}
\end{equation}
The QFI tensor becomes
\begin{align}
&\mathcal F_{ab}(\mathbf p)
    =
    \mathcal F^{\rm cla}_{ab}(\mathbf p)+\mathcal F^{\rm coh}_{ab}(\mathbf p)\nonumber\\
    =&
    \sum_{n=\pm}
    \frac{
    (\partial_a\lambda_n)(\partial_b\lambda_n)
    }{
    \lambda_n
    }
    +
    \frac{
        2|\mathbf d|^2
    }{
        d_0(d_0^2-|\mathbf d|^2)
    }
    \,
    \partial_a\hat{\mathbf d}
    \cdot
    \partial_b\hat{\mathbf d},    
     \label{eq:fab}
\end{align}
where $\hat{\mathbf d}=\mathbf d/|\mathbf d|$, and the two eigenvalues of the pair propagator are \(\lambda_\pm(\mathbf p)=\left[d_0(\mathbf p)\pm |\mathbf d(\mathbf p)|\right]^{-1}\). Stability in the Gaussian normal state requires $d_0(\mathbf p)\geq|\mathbf d(\mathbf p)|$. Therefore,
\begin{equation}
    2|\mathcal U_{xy}(\mathbf p)|=\frac{|\mathbf d(\mathbf p)|}{d_0(\mathbf p)}\sqrt{
        \det\mathcal F^{\rm coh}_{xy}}\leq\sqrt{
        \det\mathcal F^{\rm coh}_{xy}}.
    \label{eq:integrated_geometric_hierarchy}  
\end{equation}
The pointwise two-band bound approaches equality near the critical point
where the soft branch satisfies \(d_0\simeq |d|\).

\section*{Supplementary Note III: Dimensionless integrals and critical asymptotics}

In this section, we derive the dimensionless integrals used in the \textbf{Main Text} and extract their leading singularities as \(r\to0^+\). We use the
same notation as in the \textbf{Main Text} and define
\(x\equiv \gamma_0p^2/|\delta a|\). Here \(x\) is the dimensionless radial momentum variable. The inverse pair propagator is
\begin{equation}
    \hat L_0^{-1}(\mathbf p)
    =
    \begin{pmatrix}
        a_+ + \gamma_0 p^2
        &
        \gamma_m p_+^2+\lambda p_-
        \\
        \gamma_m^* p_-^2+\lambda^* p_+
        &
        a_-+\gamma_0p^2
    \end{pmatrix},
    \label{eq:SM_minimal_kernel}
\end{equation}
where \(p_\pm=p_x\pm i p_y\).  It is useful to define two dimensionless quantities from Eq.~\eqref{eq:SM_minimal_kernel} as
\begin{equation}
    \alpha(x)=\frac{d_0}{|\delta a|}=1+r+x, \qquad q(x,\theta)\equiv \frac{|\mathbf d|}{|\delta a|}.
    \label{eq:SM_ad_def}
\end{equation}
Explicitly,
\begin{equation}
\begin{split}
    q^2(x,\theta)
    &=1+\ell^2 x+g^2x^2
    +2g\ell x^{3/2}\cos(3\theta+\varphi),
\end{split}
    \label{eq:SM_q_def}
\end{equation}
where \(\varphi=\arg(\gamma_m\lambda^*)\) fixes the orientation of the
threefold anisotropy.  The two eigenvalues of the inverse propagator are
\(|\delta a|[\alpha(x)\pm q(x,\theta)]\), so that the pair-propagator
eigenvalues are
\begin{equation}
    \lambda_\pm(\mathbf p)
    =\frac{1}{|\delta a|[\alpha(x)\pm q(x,\theta)]} .
\end{equation}
At the transition, the branch \(\alpha-q\) becomes soft at \(x=0\), while the
branch \(\alpha+q\) remains massive.

\subsection*{A. Geometric factors and dimensionless response functions}

Because the system has \(C_{3\perp}\) symmetry, the longitudinal response
is governed by the rotationally averaged component
\begin{equation}
    \mathcal F_L
    =
    \frac{1}{2}
    (\mathcal F_{xx}+\mathcal F_{yy}).
\end{equation}
We define dimensionless response factors by
\begin{equation}
    \mathcal F_L^{\rm cla}(\mathbf p)
    =
    \frac{\gamma_0}{|\delta a|^2}
    f_{\rm cla}(x,\theta;r),
    \qquad
    \mathcal F_L^{\rm coh}(\mathbf p)
    =
    \frac{\gamma_0}{|\delta a|^2}
    f_{\rm coh}(x,\theta;r),
    \label{eq:SM_F_dimensionless}
\end{equation}
and
\begin{equation}
    \mathcal U_{xy}(\mathbf p)
    =
    \frac{\gamma_0\delta a}{|\delta a|^3}
    f_H(x,\theta;r).
    \label{eq:SM_U_dimensionless}
\end{equation}
The dimensionless classical longitudinal factor is
\begin{equation}
    f_{\rm cla}
    =
    \frac{1}{2}
    \sum_{s=\pm}
    \frac{
        |\nabla_{\mathbf k}[\alpha(x)+s q(x,\theta)]|^2
    }{
        [\alpha(x)+s q(x,\theta)]^3
    },
    \label{eq:SM_fcla}
\end{equation}
where \(\mathbf k\equiv \sqrt{\gamma_0/|\delta a|}\,\mathbf p\) and
\(x=k^2\). The coherent longitudinal factor is
\begin{equation}
    f_{\rm coh}
    =
    \frac{
        2\ell^2+8g^2x-
        \mathcal V(x,\theta)/q^2(x,\theta)
    }{
        \alpha(x)[\alpha^2(x)-q^2(x,\theta)]
    },
    \label{eq:SM_fcoh}
\end{equation}
with
\begin{equation}
\begin{split}
    \mathcal V(x,\theta)
    =&
    \left(
        \ell^2\sqrt{x}
        +2g^2x^{3/2}
        +3g\ell x\cos(3\theta+\varphi)
    \right)^2
    \\
    &+9g^2\ell^2x^2\sin^2(3\theta+\varphi).
\end{split}
    \label{eq:SM_V_def}
\end{equation}
The Hall factor is
\begin{equation}
    f_H
    =
    \frac{
        \ell^2-4g^2x
    }{
        \alpha^2(x)[\alpha^2(x)-q^2(x,\theta)]
    }.
    \label{eq:SM_fH}
\end{equation}
This expression already shows the physical origin of the Hall response:
the constant term in the numerator is generated by the Lifshitz invariant \(\lambda p_-\),
whereas the quadratic mixing \(\gamma_m p_+^2\) contributes through a
term proportional to \(p^2\).

In the weakly reactive limit, the longitudinal paraconductivity is
\begin{equation}
    \sigma_L^{\rm cla/coh}
    =
    \frac{e^{*2}T\Xi'}{2}
    \int\frac{d^2p}{(2\pi)^2}
    \mathcal F_L^{\rm cla/coh}(\mathbf p).
    \label{eq:SM_sigma_L_start}
\end{equation}
Using
\begin{equation}
    \frac{d^2p}{(2\pi)^2}
    =
    \frac{|\delta a|}{2\gamma_0}
    \frac{dx\,d\theta}{(2\pi)^2},
\end{equation}
we obtain
\begin{equation}
    \sigma_L^{\rm cla/coh}
    =
    \frac{e^{*2}T\Xi'}{4\pi|\delta a|}
    H_{\rm cla/coh}(r),
    \label{eq:SM_sigma_L_H}
\end{equation}
where the dimensionless functions are
\begin{equation}
    H_{\rm cla}(r)
    =
    \frac{1}{4\pi}
    \int_0^\infty dx\int_0^{2\pi}d\theta\,
    f_{\rm cla}(x,\theta;r),
    \label{eq:SM_Hcla_def}
\end{equation}
and
\begin{equation}
    H_{\rm coh}(r)
    =
    \frac{1}{4\pi}
    \int_0^\infty dx\int_0^{2\pi}d\theta\,
    f_{\rm coh}(x,\theta;r).
    \label{eq:SM_Hcoh_def}
\end{equation}
Similarly, the intrinsic fluctuational anomalous Hall conductivity is
\begin{equation}
    \sigma_H^{\rm para}
    =
    e^{*2}T\Xi''
    \int\frac{d^2p}{(2\pi)^2}\,
    \mathcal U_{xy}(\mathbf p)
    =
    \frac{e^{*2}T\Xi''}{4\pi}
    \frac{\delta a}{|\delta a|^2}
    I(r),
    \label{eq:SM_sigma_H_I}
\end{equation}
where
\begin{equation}
    I(r)
    =
    \frac{1}{2\pi}
    \int_0^\infty dx\int_0^{2\pi}d\theta\,
    f_H(x,\theta;r).
    \label{eq:SM_I_def}
\end{equation}

For the Hall response, the angular integral can be performed explicitly.
From Eqs.~\eqref{eq:SM_ad_def} and \eqref{eq:SM_q_def},
\begin{equation}
    \alpha^2(x)-q^2(x,\theta)
    =
    \mathcal D_r(x)
    -2g\ell x^{3/2}\cos(3\theta+\varphi),
\end{equation}
where
\begin{equation}
    \mathcal D_r(x)
    =
    2r+r^2+(2+2r-\ell^2)x+(1-g^2)x^2 .
    \label{eq:SM_D_def}
\end{equation}
Using
\begin{equation}
    \frac{1}{2\pi}
    \int_0^{2\pi}
    \frac{d\theta}{
        \mathcal D_r(x)-2g\ell x^{3/2}\cos(3\theta+\varphi)
    }
    =
    \frac{1}{
    \sqrt{\mathcal D_r^2(x)-4g^2\ell^2x^3}}
\end{equation}
whenever the Gaussian kernel is positive, Eq.~\eqref{eq:SM_I_def}
becomes
\begin{equation}
    I(r)
    =
    \int_0^\infty dx\,
    \frac{\ell^2-4g^2x}{
        (1+r+x)^2
        \sqrt{\mathcal D_r^2(x)-4g^2\ell^2x^3}
    } .
    \label{eq:SM_I_angle_integrated}
\end{equation}
This one-dimensional representation is the most convenient form for
numerical evaluation.  It is also useful for extracting the logarithmic
singularity analytically, as shown below.

\subsection*{B. Infrared analysis of the classical longitudinal sector}

Before evaluating the critical behavior, we identify the momentum region responsible for the singularity as the infrared
region \(x\sim r\ll1\), because the transition is
reached when the lower eigenvalue of the inverse pair propagator becomes
soft at \(\mathbf p=0\). The singular behavior near \(T_c\) is controlled entirely by the soft
branch
\begin{equation}
    A_-(x,\theta)=\alpha(x)-q(x,\theta).
\end{equation}
For \(x\ll1\), Eq.~\eqref{eq:SM_q_def} gives
\begin{equation}
    q(x,\theta)
    =
    1+\frac{\ell^2}{2}x
    +g\ell x^{3/2}\cos(3\theta+\varphi)
    +\mathcal O(x^2).
\end{equation}
Therefore,
\begin{equation}
    A_-(x,\theta)
    =
    r+\kappa_\ell x
    -g\ell x^{3/2}\cos(3\theta+\varphi)
    +\mathcal O(x^2),
    \label{eq:SM_soft_mass_expansion}
\end{equation}
where \( \kappa_\ell
    =1-\ell^2/2\). The condition \(\ell^2<2(1-g^2)\), used in the \textbf{Main Text}, guarantees the absence of an incommensurate finite-momentum instability in the full Gaussian kernel. For the infrared asymptotics derived below, the essential requirement is
\(\kappa_\ell>0\). 

The classical longitudinal factor contains derivatives of the soft eigenvalue.
Keeping only the soft branch and using \(x=k^2\), we have
\begin{equation}
    f_{\rm cla}(x,\theta;r)
    \simeq
    \frac{1}{2}
    \frac{|\nabla_{\mathbf k}A_-(x,\theta)|^2}{A_-^3(x,\theta)}.
\end{equation}
The threefold-anisotropic term in Eq.~\eqref{eq:SM_soft_mass_expansion}
is of order \(x^{3/2}\).  It is subleading for the leading infrared
singularity, so we first keep only
\begin{equation}
    A_-(x,\theta)\simeq r+\kappa_\ell x.
\end{equation}
Then
\begin{equation}
    |\nabla_{\mathbf k}A_-|^2
    =
    |2\kappa_\ell\mathbf k|^2
    =4\kappa_\ell^2x,
\end{equation}
and hence
\begin{equation}
    f_{\rm cla}(x,
    \theta;r)
    \simeq
    \frac{2\kappa_\ell^2x}{(r+\kappa_\ell x)^3}.
    \label{eq:SM_fcla_asym}
\end{equation}
Substituting this into Eq.~\eqref{eq:SM_Hcla_def} gives
\begin{equation}
\begin{split}
    H_{\rm cla}(r)
    &\simeq
    \frac{1}{4\pi}
    \int_0^{2\pi}d\theta
    \int_0^\infty dx\,
    \frac{2\kappa_\ell^2x}{(r+\kappa_\ell x)^3}
    \\
    &=
    \frac{1}{2}
    \int_0^\infty dx\,
    \frac{2\kappa_\ell^2x}{(r+\kappa_\ell x)^3}.
\end{split}
    \label{eq:SM_Hcla_intermediate}
\end{equation}
The remaining integral is elementary.  Let \(u=r+\kappa_\ell x\).  Then
\(x=(u-r)/\kappa_\ell\), \(dx=du/\kappa_\ell\), and
\begin{equation}
\begin{split}
    \int_0^\infty dx\,
    \frac{2\kappa_\ell^2x}{(r+\kappa_\ell x)^3}
    &=
    \int_r^\infty du\,
    \frac{2(u-r)}{u^3}=
    \frac{1}{r}.
\end{split}
\end{equation}
Thus
\begin{equation}
    H_{\rm cla}(r)
    \rightarrow
    \frac{1}{2r},
    \qquad
    \sigma_L^{\rm cla}
    \rightarrow
    \frac{e^{*2}T_c\Xi'}{8\pi|\delta a|}
    \frac{1}{r}.
    \label{eq:SM_Hcla_asym}
\end{equation}
The coefficient is independent of \(\ell\) because the \(\kappa_\ell\)
dependence cancels after the change of variables.  Physically, this is
the usual two-dimensional Aslamazov--Larkin divergence of the spectral
fluctuation channel.

\subsection*{C. Infrared analysis of the coherent longitudinal sector}

Having established that the critical behavior is controlled by the
infrared soft denominator \(r+\kappa_\ell x\), we now analyze the
coherent part of the QFI.  The denominator in
Eq.~\eqref{eq:SM_fcoh} is
\begin{equation}
    \alpha(x)[\alpha^2(x)-q^2(x,\theta)].
\end{equation}
As \(r\to0^+\), only the factor \(\alpha-q\) becomes soft at the transition; the other two factors, \(\alpha\simeq1\) and \(\alpha+q\simeq2\), remain massive and only contribute a finite prefactor. Thus we obtain
\begin{equation}
    \alpha(x)[\alpha^2(x)-q^2(x,\theta)]
    =
    \alpha(x)[\alpha(x)-q(x,\theta)][\alpha(x)+q(x,\theta)]
    \simeq
    2(r+\kappa_\ell x).
    \label{eq:SM_denominator_asym}
\end{equation}
The numerator in Eq.~\eqref{eq:SM_fcoh} has a finite limit at the soft
point:
\begin{equation}
    2\ell^2+8g^2x-\frac{\mathcal V(x,\theta)}{q^2(x,\theta)}
    =
    2\ell^2+\mathcal O(x).
\end{equation}
Therefore the leading coherent factor is
\begin{equation}
    f_{\rm coh}(x,\theta;r)
    \simeq
    \frac{\ell^2}{r+\kappa_\ell x}.
    \label{eq:SM_fcoh_asym}
\end{equation}
Only the infrared region is responsible for the singularity. To extract the singularity explicitly, introduce an intermediate cutoff
\(x_\Lambda\) satisfying \(r\ll x_\Lambda\ll1\). The exact integral still extends to infinity. The integral over \(0<x<x_\Lambda\) gives the singular part, while the
integral over \(x>x_\Lambda\) remains finite as \(r\to0\) and only
contributes to the non-universal \(\mathcal O(1)\) background. We therefore integrate Eq.~\eqref{eq:SM_fcoh_asym} up to this cutoff \(x_\Lambda\):
\begin{equation}
\begin{split}
    H_{\rm coh}(r)
    &\simeq
    \frac{1}{4\pi}
    \int_0^{2\pi}d\theta
    \int_0^{x_\Lambda}dx\,
    \frac{\ell^2}{r+\kappa_\ell x}
    \\
    &=
    \frac{\ell^2}{2}
    \int_0^{x_\Lambda}
    \frac{dx}{r+\kappa_\ell x}
    =
    \frac{\ell^2}{2\kappa_\ell}
    \log\frac{r+\kappa_\ell x_\Lambda}{r}.
\end{split}
\end{equation}
As \(r\to0^+\), this yields
\begin{equation}
    H_{\rm coh}(r)
    \rightarrow
    \frac{\ell^2}{2\kappa_\ell}
    \log\frac{1}{r}+\mathcal O(1).
    \label{eq:SM_Hcoh_asym}
\end{equation}
The coherent contribution is weaker than the classical contribution
because it contains only one soft propagator.  It is nevertheless
singular in two dimensions because the Lifshitz invariant makes the
pairing eigenvector vary linearly with momentum, so the quantum metric
of the soft-mode eigenvector remains finite at \(\mathbf p=0\).

\subsection*{D. Infrared analysis of the Hall sector}

The Hall sector can be analyzed from the angular-integrated form
\eqref{eq:SM_I_angle_integrated}. In the infrared regime
\(r\ll1\), \(x\ll1\), the function \(\mathcal D_r(x)\) becomes
\begin{equation}
    \mathcal D_r(x)
    =
    2r+(2-\ell^2)x
    +\mathcal O(r^2,rx,x^2)
    =
    2(r+\kappa_\ell x)
    +\mathcal O(r^2,rx,x^2).
    \label{eq:SM_Dr_IR}
\end{equation}
Therefore the square-root denominator is
\begin{equation}
\begin{split}
    \sqrt{
        \mathcal D_r^2(x)-4g^2\ell^2x^3
    }
    &=
    \sqrt{
        4(r+\kappa_\ell x)^2
        -4g^2\ell^2x^3
        +\cdots
    } .
\end{split}
    \label{eq:SM_sqrt_before_expand}
\end{equation}
The second term inside the square root is subleading in the infrared
region. Thus,
\begin{equation}
    \sqrt{
        \mathcal D_r^2(x)-4g^2\ell^2x^3
    }
    =
    2(r+\kappa_\ell x)
    \left[1+\mathcal O(x)+\mathcal O(r)\right].
    \label{eq:SM_sqrt_IR}
\end{equation}
Similarly, the remaining factors in the integrand have the infrared expansions. The numerator becomes
\begin{equation}
    \ell^2-4g^2x
    =
    \ell^2+\mathcal O(x),
\end{equation}
and in the denominator
\begin{equation}
    (1+r+x)^2
    =
    1+\mathcal O(r,x).
\end{equation}
Substituting these results into
Eq.~\eqref{eq:SM_I_angle_integrated}, the singular part of the
integrand is
\begin{equation}
    \frac{
        \ell^2-4g^2x
    }{
        (1+r+x)^2
        \sqrt{
        \mathcal D_r^2(x)-4g^2\ell^2x^3
        }
    }
    =
    \frac{
        \ell^2
    }{
        2(r+\kappa_\ell x)
    }
    +
    \text{regular terms}.
    \label{eq:SM_Ir_integrand_asym}
\end{equation}
This equation is the key infrared result.  It shows that the Hall
integral has the same soft denominator \(r+\kappa_\ell x\) as the
coherent longitudinal sector.  Since \(x\propto p^2\), this is the
radial version of the two-dimensional infrared integral
\(\int d^2p/(r+p^2)\). Following the same infrared analysis as for $H_{\rm coh}(r)$,
\begin{equation}
\begin{split}
    I(r)
    &\simeq
    \frac{\ell^2}{2}
    \int_0^{x_\Lambda}
    \frac{dx}{r+\kappa_\ell x}
   =
    \frac{\ell^2}{2\kappa_\ell}
    \log
    \frac{
        r+\kappa_\ell x_\Lambda
    }{
        r
    } .
\end{split}
    \label{eq:SM_Ir_sing_integral}
\end{equation}
As \(r\to0^+\), this becomes
\begin{equation}
    I(r)
    \rightarrow
    \frac{
        \ell^2
    }{
        2\kappa_\ell
    }
    \log\frac{1}{r}
    +\mathcal O(1).
    \label{eq:SM_Ir_asym}
\end{equation}
This is the logarithmic divergence quoted in the \textbf{Main Text}. The logarithmic coefficient is proportional to \(\ell^2\), showing that
the Lifshitz invariant is responsible for the Hall singularity.  If
\(\ell=0\), the numerator starts as \(-4g^2x\), and the infrared integral is regular. The quadratic mixing \(g\) enters only through subleading terms in the
infrared expansion, either as \(-4g^2x\) in the numerator or as
\(4g^2\ell^2x^3\) under the square root.  It therefore modifies only the
regular \(\mathcal O(1)\) part of \(I(r)\), not the leading logarithm.


\begin{thebibliography}{10}

\bibitem{gao2025quantum}
Anyuan Gao, Naoto Nagaosa, Ni~Ni, and Su-Yang Xu.
\newblock Quantum geometry phenomena in condensed matter systems.
\newblock {\em arXiv preprint arXiv:2508.00469}, 2025.

\bibitem{yu2025quantum}
Jiabin Yu, B~Andrei Bernevig, Raquel Queiroz, Enrico Rossi, P{\"a}ivi T{\"o}rm{\"a}, and Bohm-Jung Yang.
\newblock Quantum geometry in quantum materials.
\newblock {\em npj Quantum Materials}, 10(1):101, 2025.

\bibitem{jiang2025revealing}
Yiyang Jiang, Tobias Holder, and Binghai Yan.
\newblock Revealing quantum geometry in nonlinear quantum materials.
\newblock {\em Reports on Progress in Physics}, 88(7):076502, 2025.

\bibitem{liu2025quantum}
Tianyu Liu, Xiao-Bin Qiang, Hai-Zhou Lu, and XC~Xie.
\newblock Quantum geometry in condensed matter.
\newblock {\em National Science Review}, 12(3):nwae334, 2025.

\bibitem{nagaosa2010anomalous}
Naoto Nagaosa, Jairo Sinova, Shigeki Onoda, Allan~H MacDonald, and Nai~Phuan Ong.
\newblock Anomalous hall effect.
\newblock {\em Reviews of modern physics}, 82(2):1539--1592, 2010.

\bibitem{xiao2010berry}
Di~Xiao, Ming-Che Chang, and Qian Niu.
\newblock Berry phase effects on electronic properties.
\newblock {\em Reviews of modern physics}, 82(3):1959--2007, 2010.

\bibitem{chang2023colloquium}
Cui-Zu Chang, Chao-Xing Liu, and Allan~H MacDonald.
\newblock Colloquium: quantum anomalous hall effect.
\newblock {\em Reviews of Modern Physics}, 95(1):011002, 2023.

\bibitem{vanderbilt2018berry}
David Vanderbilt.
\newblock {\em Berry phases in electronic structure theory: electric polarization, orbital magnetization and topological insulators}.
\newblock Cambridge University Press, 2018.

\bibitem{sodemann2015quantum}
Inti Sodemann and Liang Fu.
\newblock Quantum nonlinear hall effect induced by berry curvature dipole in time-reversal invariant materials.
\newblock {\em Physical review letters}, 115(21):216806, 2015.

\bibitem{gao2014field}
Yang Gao, Shengyuan~A Yang, and Qian Niu.
\newblock Field induced positional shift of bloch electrons and its dynamical implications.
\newblock {\em Physical review letters}, 112(16):166601, 2014.

\bibitem{wang2021intrinsic}
Chong Wang, Yang Gao, and Di~Xiao.
\newblock Intrinsic nonlinear hall effect in antiferromagnetic tetragonal cumnas.
\newblock {\em Physical Review Letters}, 127(27):277201, 2021.

\bibitem{liu2021intrinsic}
Huiying Liu, Jianzhou Zhao, Yue-Xin Huang, Weikang Wu, Xian-Lei Sheng, Cong Xiao, and Shengyuan~A Yang.
\newblock Intrinsic second-order anomalous hall effect and its application in compensated antiferromagnets.
\newblock {\em Physical Review Letters}, 127(27):277202, 2021.

\bibitem{zhang2023higher}
Cheng-Ping Zhang, Xue-Jian Gao, Ying-Ming Xie, and Hoi~Chun Po.
\newblock Higher-order nonlinear anomalous hall effects induced by berry curvature multipoles.
\newblock {\em Physical Review B}, 107(11):115142, 2023.

\bibitem{ahn2022riemannian}
Junyeong Ahn, Guang-Yu Guo, Naoto Nagaosa, and Ashvin Vishwanath.
\newblock Riemannian geometry of resonant optical responses.
\newblock {\em Nature Physics}, 18(3):290--295, 2022.

\bibitem{peotta2015superfluidity}
Sebastiano Peotta and P{\"a}ivi T{\"o}rm{\"a}.
\newblock Superfluidity in topologically nontrivial flat bands.
\newblock {\em Nature communications}, 6(1):8944, 2015.

\bibitem{uhlmann1986parallel}
Armin Uhlmann.
\newblock Parallel transport and “quantum holonomy” along density operators.
\newblock {\em Reports on Mathematical Physics}, 24(2):229--240, 1986.

\bibitem{uhlmann1991gauge}
Armin Uhlmann.
\newblock A gauge field governing parallel transport along mixed states.
\newblock {\em letters in mathematical physics}, 21(3):229--236, 1991.

\bibitem{braunstein1994statistical}
Samuel~L Braunstein and Carlton~M Caves.
\newblock Statistical distance and the geometry of quantum states.
\newblock {\em Physical Review Letters}, 72(22):3439, 1994.

\bibitem{carollo2018uhlmann}
Angelo Carollo, Bernardo Spagnolo, and Davide Valenti.
\newblock Uhlmann curvature in dissipative phase transitions.
\newblock {\em Scientific reports}, 8(1):9852, 2018.

\bibitem{carollo2020geometry}
Angelo Carollo, Davide Valenti, and Bernardo Spagnolo.
\newblock Geometry of quantum phase transitions.
\newblock {\em Physics Reports}, 838:1--72, 2020.

\bibitem{albert2016geometry}
Victor~V. Albert, Barry Bradlyn, Martin Fraas, and Liang Jiang.
\newblock Geometry and response of lindbladians.
\newblock {\em Physical Review X}, 6:041031, 2016.

\bibitem{leonforte2019uhlmann}
Luca Leonforte, Davide Valenti, Bernardo Spagnolo, and Angelo Carollo.
\newblock Uhlmann number in translational invariant systems.
\newblock {\em Scientific reports}, 9(1):9106, 2019.

\bibitem{liu2020quantum}
Jing Liu, Haidong Yuan, Xiao-Ming Lu, and Xiaoguang Wang.
\newblock Quantum fisher information matrix and multiparameter estimation.
\newblock {\em Journal of Physics A: Mathematical and Theoretical}, 53(2):023001, 2020.

\bibitem{ji2025density}
Guangyue Ji, David~E Palomino, Nathan Goldman, Tomoki Ozawa, Peter Riseborough, Jie Wang, and Bruno Mera.
\newblock Density matrix geometry and sum rules.
\newblock {\em arXiv preprint arXiv:2507.14028}, 2025.

\bibitem{schmid1969diamagnetic}
Albert Schmid.
\newblock Diamagnetic susceptibility at the transition to the superconducting state.
\newblock {\em Physical Review}, 180(2):527, 1969.

\bibitem{larkin2005theory}
Anatoly Larkin and Andrei Varlamov.
\newblock {\em Theory of fluctuations in superconductors}, volume 127.
\newblock OUP Oxford, 2005.

\bibitem{fukuyama1971fluctuation}
Hidetoshi Fukuyama, Hiromichi Ebisawa, and Toshio Tsuzuki.
\newblock Fluctuation of the order parameter and hall effect.
\newblock {\em Progress of Theoretical Physics}, 46(4):1028--1041, 1971.

\bibitem{aronov1995gauge}
AG~Aronov, S~Hikami, and AI~Larkin.
\newblock Gauge invariance and transport properties in superconductors above t c.
\newblock {\em Physical Review B}, 51(6):3880, 1995.

\bibitem{michaeli2012hall}
Karen Michaeli, Konstantin~S Tikhonov, and Alexander~M Finkel'Stein.
\newblock Hall effect in superconducting films.
\newblock {\em Physical Review B—Condensed Matter and Materials Physics}, 86(1):014515, 2012.

\bibitem{sumiyoshi2014giant}
Hiroaki Sumiyoshi and Satoshi Fujimoto.
\newblock Giant nernst and hall effects due to chiral superconducting fluctuations.
\newblock {\em Physical Review B}, 90(18):184518, 2014.

\bibitem{li2020fluctuational}
Songci Li and Alex Levchenko.
\newblock Fluctuational anomalous hall and nernst effects in superconductors.
\newblock {\em Annals of Physics}, 417:168137, 2020.

\bibitem{han2025signatures}
Tonghang Han, Zhengguang Lu, Zach Hadjri, Lihan Shi, Zhenghan Wu, Wei Xu, Yuxuan Yao, Armel~A Cotten, Omid Sharifi~Sedeh, Henok Weldeyesus, et~al.
\newblock Signatures of chiral superconductivity in rhombohedral graphene.
\newblock {\em Nature}, 643(8072):654--661, 2025.

\bibitem{dutta2026reconfigurable}
Surajit Dutta, Nadav Auerbach, Tonghang Han, Yaozhang Zhou, Gal Shavit, Niladri-Sekhar Kander, Yuri Myasoedov, Martin~E Huber, Kenji Watanabe, Takashi Taniguchi, et~al.
\newblock Reconfigurable chiral superconductivity.
\newblock {\em arXiv preprint arXiv:2605.13303}, 2026.

\bibitem{sheekey2026visualizing}
Owen~I Sheekey, Trevor~B Arp, Benjamin~A Foutty, Ruoxi Zhang, Tixuan Tan, Ludwig~FW Holleis, Yi~Guo, Sandesh~S Kalantre, Canxun Zhang, Mark Zakharyan, et~al.
\newblock Visualizing orbital magnetism in electron doped rhombohedral multilayer graphene.
\newblock {\em arXiv preprint arXiv:2605.30316}, 2026.

\bibitem{nagaosa2013quantum}
Naoto Nagaosa.
\newblock {\em Quantum field theory in condensed matter physics}.
\newblock Springer Science \& Business Media, 2013.

\bibitem{Altland_Simons_2023}
Alexander Altland and Ben Simons.
\newblock {\em Condensed Matter Field Theory}.
\newblock Cambridge University Press, 3 edition, 2023.

\bibitem{wu2021nature}
Xianxin Wu, Tilman Schwemmer, Tobias M{\"u}ller, Armando Consiglio, Giorgio Sangiovanni, Domenico Di~Sante, Yasir Iqbal, Werner Hanke, Andreas~P Schnyder, M~Michael Denner, et~al.
\newblock Nature of unconventional pairing in the kagome superconductors av 3 sb 5 (a= k, rb, cs).
\newblock {\em Physical review letters}, 127(17):177001, 2021.

\bibitem{kallin2016chiral}
Catherine Kallin and John Berlinsky.
\newblock Chiral superconductors.
\newblock {\em Reports on Progress in Physics}, 79(5):054502, 2016.

\bibitem{guan2026exploring}
Yuntao Guan and Barry Bradlyn.
\newblock Exploring many-body quantum geometry beyond the quantum metric with correlation functions: A time-dependent perspective.
\newblock {\em Physical Review Research}, 8(1):013291, 2026.

\bibitem{ghosh2026journey}
Priya Ghosh, Tanoy~Kanti Konar, Debraj Rakshit, Aditi~Sen De, and Ujjwal Sen.
\newblock Journey in quantum metrology and sensing from foundations to applications: a review.
\newblock {\em arXiv preprint arXiv:2605.21702}, 2026.

\bibitem{van2022bures}
Jesse van Oostrum.
\newblock Bures--wasserstein geometry for positive-definite hermitian matrices and their trace-one subset.
\newblock {\em Information Geometry}, 5(2):405--425, 2022.

\bibitem{montenegro2025quantum}
Victor Montenegro, Chiranjib Mukhopadhyay, Rozhin Yousefjani, Saubhik Sarkar, Utkarsh Mishra, Matteo~GA Paris, and Abolfazl Bayat.
\newblock Quantum metrology and sensing with many-body systems.
\newblock {\em Physics Reports}, 1134:1--62, 2025.

\bibitem{ussishkin2002gaussian}
Iddo Ussishkin, Shivaji~Lal Sondhi, and David~A Huse.
\newblock Gaussian superconducting fluctuations, thermal transport, and the nernst effect.
\newblock {\em Physical review letters}, 89(28):287001, 2002.

\bibitem{SM}
The Supplementary Materials for ``Quantum Information Geometry of Multicomponent Superconducting Fluctuation Transport", which includes three parts: I. Derivation of the kinetic equation for the pairing fluctuation density matrix, II. Quantum information geometric bounds on paraconductivity, III. Dimensionless integrals and critical asymptotics.

\bibitem{wakatsuki2017nonreciprocal}
Ryohei Wakatsuki, Yu~Saito, Shintaro Hoshino, Yuki~M Itahashi, Toshiya Ideue, Motohiko Ezawa, Yoshihiro Iwasa, and Naoto Nagaosa.
\newblock Nonreciprocal charge transport in noncentrosymmetric superconductors.
\newblock {\em Science advances}, 3(4):e1602390, 2017.

\bibitem{mishonov2003kinetics}
Todor~M Mishonov, Georgi~V Pachov, Ivan~N Genchev, Liliya~A Atanasova, and Damian~Ch Damianov.
\newblock Kinetics and boltzmann kinetic equation for fluctuation cooper pairs.
\newblock {\em Physical Review B}, 68(5):054525, 2003.

\bibitem{mandal2024quantum}
Debottam Mandal, Sanjay Sarkar, Kamal Das, and Amit Agarwal.
\newblock Quantum geometry induced third-order nonlinear transport responses.
\newblock {\em Physical Review B}, 110(19):195131, 2024.

\bibitem{daido2024rectification}
Akito Daido and Youichi Yanase.
\newblock Rectification and nonlinear hall effect by fluctuating finite-momentum cooper pairs.
\newblock {\em Physical Review Research}, 6(2):L022009, 2024.

\bibitem{ebisawa1971wave}
Hiromichi Ebisawa and Hidetoshi Fukuyama.
\newblock Wave character of the time dependent ginzburg landau equation and the fluctuating pair propagator in superconductors.
\newblock {\em Progress of Theoretical Physics}, 46(4):1042--1053, 1971.

\bibitem{ozawa2021relations}
Tomoki Ozawa and Bruno Mera.
\newblock Relations between topology and the quantum metric for chern insulators.
\newblock {\em Physical Review B}, 104(4):045103, 2021.

\bibitem{onishi2024fundamental}
Yugo Onishi and Liang Fu.
\newblock Fundamental bound on topological gap.
\newblock {\em Physical Review X}, 14(1):011052, 2024.

\bibitem{shinada2025quantum}
Koki Shinada and Naoto Nagaosa.
\newblock Quantum geometric bounds for observables: Linear responses, drude weight, and orbital magnetization.
\newblock {\em Physical Review B}, 112(15):155158, 2025.

\bibitem{geier2025chiral}
Max Geier, Margarita Davydova, and Liang Fu.
\newblock Chiral and topological superconductivity in isospin polarized multilayer graphene.
\newblock {\em Nature Communications}, 2025.

\bibitem{zhu2026microscopic}
Jihang Zhu and Chunli Huang.
\newblock Microscopic origin of orbital magnetization in chiral superconductors.
\newblock {\em arXiv preprint arXiv:2601.12387}, 2026.

\bibitem{yoon2025quarter}
Chiho Yoon, Tianyi Xu, Yafis Barlas, and Fan Zhang.
\newblock Quarter metal superconductivity.
\newblock {\em arXiv preprint arXiv:2502.17555}, 2025.

\bibitem{nagashima2026type}
Raigo Nagashima, Chihiro Mamiya, and Naoto Tsuji.
\newblock Type ii lifshitz invariant and optically active higgs mode in time-reversal symmetry broken superconductors.
\newblock {\em arXiv preprint arXiv:2604.15054}, 2026.

\bibitem{lundemo2026fluctuation}
Sondre~Duna Lundemo and Asle Sudb{\o}.
\newblock Fluctuation conductivity in ultraclean multicomponent superconductors.
\newblock {\em Physical Review B}, 113(18):184504, 2026.

\bibitem{qin2026chiral}
Qiong Qin and Congjun Wu.
\newblock Chiral finite-momentum superconductivity in the tetralayer graphene.
\newblock {\em Chinese Physics Letters}.

\bibitem{efimkin2021topological}
Dmitry~K Efimkin.
\newblock Topological fluctuating electron-hole cooper pairs in graphene-gaas heterostructures.
\newblock {\em Physical Review B}, 104(24):245436, 2021.

\bibitem{may2026pairing}
Julian May-Mann, Tobias Helbig, and Trithep Devakul.
\newblock How pairing mechanism dictates topology in valley-polarized superconductors with berry curvature.
\newblock {\em npj Quantum Materials}, 2026.

\bibitem{kalantre2026fermiology}
Sandesh~S Kalantre, Ben~H Alexander, Julian May-Mann, Jonah Herzog-Arbeitman, Marisa Hocking, Qingrui Cao, Kenji Watanabe, Takashi Taniguchi, David Goldhaber-Gordon, Andrew~J Mannix, et~al.
\newblock Fermiology and the candidate chiral superconductor in rhombohedral tetralayer graphene.
\newblock {\em arXiv preprint arXiv:2606.05356}, 2026.

\bibitem{matsumoto2025reciprocal}
Tsugumi Matsumoto, Youichi Yanase, and Akito Daido.
\newblock Reciprocal and nonreciprocal paraconductivity in bilayer multiphase superconductors.
\newblock {\em Physical Review B}, 111(6):064501, 2025.

\bibitem{dong2025enhanced}
Zi-Hao Dong, Hui Yang, and Yi~Zhang.
\newblock Enhanced nonlinear hall effect by cooper pairs near the superconducting phase transition.
\newblock {\em Physical Review B}, 111(15):155120, 2025.

\bibitem{matsumoto2011rotational}
Ryo Matsumoto and Shuichi Murakami.
\newblock Rotational motion of magnons and the thermal hall effect.
\newblock {\em Physical Review B—Condensed Matter and Materials Physics}, 84(18):184406, 2011.

\bibitem{qin2012berry}
Tao Qin, Jianhui Zhou, and Junren Shi.
\newblock Berry curvature and the phonon hall effect.
\newblock {\em Physical Review B—Condensed Matter and Materials Physics}, 86(10):104305, 2012.

\bibitem{pokharel2022dynamics}
Amrit~Raj Pokharel, Vladimir Grigorev, Arjan Mejas, Tao Dong, Amir~A Haghighirad, Rolf Heid, Yi~Yao, Michael Merz, Matthieu Le~Tacon, and Jure Demsar.
\newblock Dynamics of collective modes in an unconventional charge density wave system bani2as2.
\newblock {\em Communications Physics}, 5(1):141, 2022.

\bibitem{chu2012divergent}
Jiun-Haw Chu, Hsueh-Hui Kuo, James~G Analytis, and Ian~R Fisher.
\newblock Divergent nematic susceptibility in an iron arsenide superconductor.
\newblock {\em Science}, 337(6095):710--712, 2012.

\bibitem{chen2025generalized}
Shuai~A Chen, Roderich Moessner, and Tai~Kai Ng.
\newblock Generalized peierls substitution for wannier obstructions: Response to disorder and interactions.
\newblock {\em Physical Review Letters}, 135(11):116502, 2025.

\end{thebibliography}
\end{document}